\documentclass[twocolumn]{aastex631}
\usepackage{multirow}
\usepackage[hang,flushmargin]{footmisc}
\usepackage{xspace}
\usepackage{placeins}
\usepackage{overpic}
\usepackage{tikz}
\usepackage{nth}
\usepackage[thinc]{esdiff}
\usepackage{amsopn}

\graphicspath{{./}{figures/}}
\renewcommand{\AA}{\r{A}}

\newcommand{\lya}{Ly$ \alpha $\xspace}
\newcommand{\rto}{\citetalias{richard21}\xspace}

\newcommand{\SI}[2]{#1\,\mathrm{#2}}

\newcommand{\heii}{\ion{He}{2}$ \lambda 1640 $\xspace}
\newcommand{\ciii}{\ion{C}{3}]$ \lambda\lambda 1907,1909 $\xspace}
\newcommand{\civ}{\ion{C}{4}$ \lambda\lambda 1548,1551 $\xspace}
\newcommand{\oiii}{\ion{O}{3}]$ \lambda\lambda 1661,1666 $\xspace}
\newcommand{\siiii}{\ion{Si}{3}]$ \lambda\lambda 1883,1892 $\xspace}
\newcommand{\nv}{\ion{N}{5}$ \lambda\lambda 1238,1243 $\xspace}
\newcommand{\siii}{\ion{Si}{2}$ \lambda 1260 $\xspace}
\newcommand{\siiih}{\ion{Si}{2}$ \lambda 1527 $\xspace}
\newcommand{\cii}{\ion{C}{2}$ \lambda 1334 $\xspace}
\newcommand{\siiv}{\ion{Si}{4}$ \lambda\lambda 1394,1403 $\xspace}

\newcommand{\ovi}{\ion{O}{6}$ \lambda\lambda 1032,1038 $\xspace}
\newcommand{\niv}{\ion{N}{4}]$ \lambda\lambda 1483,1487 $\xspace}
\newcommand{\niii}{\ion{N}{3}]$ \lambda 1750 $\xspace}
\newcommand{\oi}{\ion{O}{1}$ \lambda 1302 $\xspace}
\defcitealias{richard21}{R21}
\defcitealias{vanzella17}{V17}
\defcitealias{spilker20a}{S20}
\defcitealias{spilker18}{S18}

\begin{document}
	
	
	\title{Delayed Feedback in High-$ z $ Starbursts Revealed by Lyman-$ \alpha $ Profiles and Metal Line Diagnostics}
	
	\correspondingauthor{James Nianias}
	\email{nianias@hku.hk}
	
	\author[0000-0001-6985-2939]{James Nianias}
	\affiliation{Department of Physics, University of Hong Kong, Pokfulam Road, Hong Kong}
	
	\author[0000-0003-4220-2404]{Jeremy Lim}
	\affiliation{Department of Physics, University of Hong Kong, Pokfulam Road, Hong Kong}
	
	\author{Yik Lok Wong}
	\affiliation{Department of Physics and Astronomy, The University of Manchester, Oxford Road, Manchester M13 9PL, UK}
	
	\author{Gordon Wong}
	\affiliation{Department of Physics, University of Hong Kong, Pokfulam Road, Hong Kong}
	

\begin{abstract}

Lyman-$ \alpha $ emission, which couples strongly to gas kinematics, is a key observable for probing outflows from star-forming galaxies in the early universe.  Inferences of outflow properties from Lyman-$ \alpha $, however, often lack contextual comparisons with more direct outflow diagnostics from down-the-barrel metal absorption lines and driving-source properties from metal emission lines.  Here, we make such checks by taking advantage of the lensing magnification provided by galaxy clusters for 573 Lyman-$ \alpha $ sources observed with the Multi-Unit Spectroscopic Explorer (MUSE).  Using metal emission lines to measure systemic redshifts, we confirm that the Lyman-$ \alpha $ profiles are consistent with outflowing gas: single peaks redshifted relative to, or double peaks straddling, the systemic redshift.  In cases where metal absorption lines are detected, their blueshifted velocities indicate outflows, while their line ratios point to absorption by a clumpy medium.  We demonstrate a significant positive correlation between absorption-line outflow velocity and \lya line dispersion, confirming that \lya profile morphology encodes useful kinematic information about outflowing gas.  We further find that the equivalent width of high-ionization metal emission lines are anticorrelated with measured outflow velocity, suggesting younger stellar populations produce slower outflows, and providing a critical empirical constraint on feedback timescales in high-$ z $ starbursts.  Fitting model Lyman-$ \alpha $ profiles based on simple expanding shell geometry to those observed, we find that such models successfully reproduce some aspects of the data, yet often yield unphysical parameters -- calling for caution when inferring outflow properties from such models.

\end{abstract}

\section{Introduction}
\label{section:intro}

Feedback from star formation is believed to play a pivotal role in regulating galaxy evolution, particularly in the early universe where intense starbursts can drive powerful outflows that suppress further star formation and eject metals into the circumgalactic medium (CGM) and/or intergalactic medium (IGM).  The most definitive tracer of outflows in the early universe is blueshifted absorption from gas accelerated along the line of sight (``down-the-barrel" absorption), which provides unambiguous evidence of outflowing material.  Rest-frame UV absorption lines (accessible by ground-based observatories at $ z \gtrsim 2 $) have been used to diagnose outflows in stacked spectra of Lyman Break Galaxies (LBGs) \citep{shapley03,steidel10,jones12}, and Lyman-$ \alpha $ (\lya)-emitting galaxies \citep{berry12}.  However, detections of such absorption signatures in individual high-$ z $ galaxies are limited to a relatively small number of highly luminous systems (e.g. \citealt{pettini02}, \citealt{shibuya14}), which are unlikely to be representative of typical star-forming galaxies (SFGs) in the early universe.

To search for outflows in individual low-luminosity SFGs in the early universe, observers have instead turned to the \lya emission line.  Due to the abundant \lya photons produced in star forming regions via re-processing of extreme UV radiation ($ \lambda < \SI{912}{\text{\AA}} $) from young stars, even SFGs that are extremely faint in the continuum can be detected via the \lya line \citep{partridge67}.  Moreover, \lya's large scattering cross-section with neutral hydrogen (H) means that the emergent line profile is strongly coupled to the kinematics of the interstellar medium (ISM) and CGM, and hence encodes information about outflows (see reviews by \citealt{erb15} and \citealt{ouchi20}).  Most \lya-emitting galaxies display asymmetric \lya peaks that are redshifted relative to the systemic redshift of the host galaxy (where measured), sometimes accompanied by additional weaker blueshifted peaks (e.g. \citealt{kulas12}, \citealt{trainor15}, \citealt{rivera-thorsen15}).  These distinctive line profiles can be reproduced by radiative transfer (RT) models consisting of expanding shells/halos of gas (e.g. \citealt{ahn04}, \citealt{verhamme06}, \citealt{orsi12}, \citealt{duval14}), thus providing theoretical support for outflows from these objects.  Fitting such models to observed \lya profiles provides estimates of outflow parameters such as expansion velocity and neutral column density \citep{gronke15,orlitova18,gl19,gl22}, but is fraught with degeneracies and uncertainties owing to the complexity of RT in the \lya line \citep{gronke16,li22}.  Comparisons between \lya lines and down-the-barrel absorption lines -- that could be used to better understand the impact of outflow properties on the emergent \lya line profile -- are lacking, particularly in the early universe.

In this paper, we study a large sample ($ N = 573 $) of gravitationally-lensed \lya-emitting galaxies observed with the Multi-Unit Spectroscopic Explorer (MUSE; \citealt{bacon10}) and cataloged by \cite{richard21} (\rto henceforth).  Drawing on MUSE observations of 12 galaxy clusters, with accompanying deep Hubble Space Telescope (HST) observations, \rto uncovered a large number of \lya-emitting galaxies at $ 2.9 < z < 6.7 $.  The intrinsic (i.e. un-lensed) morphologies of these sources were subsequently studied in further detail by \cite{claeyssens22}.  Thanks to the magnification provided by gravitational lensing, some of these objects are among the faintest and most compact galaxies yet seen in the early universe (see, e.g., \citealt{vanzella17,mestric22,messa24,iani24}).  The boost in signal-to-noise ratio (SNR) provided by gravitational lensing also enables the detection of faint spectral features that may otherwise be undetected: UV nebular emission lines (besides \lya) and gas-phase metal absorption lines that may originate from the ISM, CGM, and/or outflows.  We analyze the spectra of these objects to search for evidence of outflows by (i) using optically-thin UV emission lines (where detected) to trace their systemic velocities, against which the \lya profiles can be compared, and (ii) searching for evidence of blueshifted metal absorption lines from outflowing gas.  We then investigate the relationship between metal absorption profiles and emergent \lya profiles, developing an empirically-driven understanding of how to interpret \lya profiles in terms of outflow properties.  Comparing \lya profiles with metal emission lines from \ion{H}{2} regions, we investigate how outflows are affected by the evolution of the stellar populations that drive them.  We complement this analysis by fitting expanding shell RT models to the \lya profiles, testing the ability of such idealized models to infer outflow properties by comparing them against the observables obtained from metal emission and absorption lines.

In Section \ref{section:obs}, we provide details about the catalogs of \rto, on which our analysis is based, as well as basic details of the MUSE observations from which the spectra are derived. We also provide details of the fitting routines and flagging procedures that we use to detect and verify emission/absorption lines.  In Section \ref{section:results}, we present the results of our fitting, which suggest that clumpy, multiphase outflows are widespread among high-$ z $ \lya-emitting galaxies. We further show -- for the first time -- significant correlations between \lya profile shape and both metal absorption-line outflow velocities and metal emission-line equivalent widths, providing direct evidence for strengthening feedback within the first $ \sim \SI{10}{Myr} $ of an episode of star formation, presumably due to the onset of supernova explosions (SNe).  We then go on to fit homogeneous expanding shell RT models to the \lya profiles and assess their success in reproducing the observed outflow properties as revealed by metal absorption lines.  We find that, while these models can generally reproduce the \lya profiles well, the fitted parameters are in many cases unphysical, which can likely be attributed to discrepancies between homogeneous thin shell models and real outflow geometry.  We discuss the implications of our results for feedback timescales in the early universe in Section \ref{section:discussion}.  Readers interested only in a concise summary of the main results can skip ahead to Section \ref{section:sum}.  Throughout this paper, we adopt a flat $\Lambda$CDM cosmology with $ H_0 = \SI{68}{km\,s^{-1}\,Mpc^{-1}}$ and $\Omega_{\text{M}} = 0.31$ \citep{planck14}.

\begin{deluxetable*}{ccccc}
	\tablecaption{Information about the MUSE cubes for each cluster.  From left to right: cluster name, cluster redshift, coordinates at center of MUSE cube, effective integration time (from \rto), sensitivity for an aperture of diameter equal to $ 2 \times $ FWHM, using the median per-pixel variance at $ \SI{6500}{\AA} $}
	\tablehead{
		\colhead{Cluster} & \colhead{$ z_{clus} $} & \colhead{Coordinates} & \colhead{$ T_{\text{eff}} $} & \colhead{$ 1\sigma $ sensitivity} \\
		\colhead{} & \colhead{} & \colhead{RA, DEC} & \colhead{hrs} & \colhead{$ \SI{10^{-20}}{erg\,s^{-1}\,cm^{-2}\,\AA^{-1}} $}
	}	 
	\startdata
	A2744 & 0.308 & 00:14:20.702, -30:24:00.63 & 3.5--7 & 12.20 \\
	A370 & 0.375 & 02:39:53.122, -01:34:56.14 & 1.5--8.5 & 11.64 \\		
	MACS0257 & 0.322 & 02:57:41.070, -22:09:17.70 & 8 & 5.63 \\ 
	MACS0329 & 0.450 & 03:29:41.568, -02:11:46.41 & 2.5 & 10.23 \\
	MACS0416(NE) & 0.397 & 04:16:09.144, -24:04:02.95 & 17 & 4.75 \\
	MACS0416(S) & 0.397 & 04:16:09.144, -24:04:02.95 & 11--15 & 8.38 \\
	BULLET & 0.296 & 06:58:38.126, -55:57:25.87 & 2 & 8.53 \\
	MACS0940 & 0.335 & 09:40:53.698, +07:44:25.31 & 8 & 6.09 \\ 
	MACS1206 & 0.438 & 12:06:12.149, -08:48:03.37 & 4--9 & 8.98 \\ 
	RXJ1347 & 0.451 & 13:47:30.617, -11:45:09.51 & 2--3 & 7.66 \\
	SMACS2031 & 0.331 & 20:31:53.256, -40:37:30.79 & 10 & 10.07 \\ 
	SMACS2131 & 0.442 & 21:31:04.831, -40:19:20.92 & 7 & 4.43 \\
	MACS2214 & 0.502 & 22:14:57.292, -14:00:12.91 & 7 & 5.93
	\enddata
\end{deluxetable*}
\label{table:cubes}

\section{Archival Observations and Data Pipeline}
\label{section:obs}

\subsection{Catalog and Sample Selection}
\label{section:catalog}

We used the spectroscopic catalogs compiled by \rto to identify high-$ z $ \lya-emitting galaxies.  Here, we provide a basic overview of how these catalogs were compiled; full details can be found in \rto.  The sources were detected either based on continuum emission in stacked HST images using \texttt{SExtractor} \citep{sextractor}, or based on line emission seen in the MUSE cubes using \texttt{MPDAF} (\textit{the MUSE Python Data Analysis Framework}, which provides a suite of tools for working with MUSE data; see \citealt{mpdafref}).  The catalogs provide source positions (flux-weighted centroids), spectroscopic redshifts, apparent magnitudes in HST ACS/WFC filters (F435W, F606W, F814W, F105W, F125W, F140W, and F160W), and estimated lensing magnifications, among other information.  Using the segmentation maps generated by \texttt{SExtractor}, \rto then extract spectra for each source.  In separate catalogs, \rto provide information about each spectral line in each spectrum including total flux, central wavelength, and full width at half-maximum (FWHM).  They estimate these parameters using the spectral fitting module \texttt{Pyplatefit}\footnote{\url{https://pyplatefit.readthedocs.io/en/latest/tutorial.html}}, which fits families of lines (or templates), such that all lines in each family are fitted simultaneously with the same central velocity and FWHM.  In emission, the resonant doublets \ovi, \nv, and \civ are each fitted as distinct families, while the forbidden \niv, \heii, \oiii, \niii, \siiii, and \ciii lines are all fitted simultaneously.  In absorption, 19 lines are fitted simultaneously, including the aforementioned resonant doublets, as well as \siii, \oi, \cii, and \siiv.  The continuum is fitted using stellar population models of \cite{bc03}.  For more information on \texttt{Pyplatefit}, see \rto.  The advantage of this approach is that it can provide accurate redshifts even in cases where the SNR of individual lines is too low for them to be detectable.  The downside is that any systematics affecting one or two lines can be carried forward to the rest of the template; this is especially concerning since MUSE spectra inevitably contain some residual atmospheric features, which can masquerade as emission/absorption lines with high SNR.  Any systematics resulting from the continuum subtraction (due to, for example, contamination by light from cluster members, or residual atmospheric continuum) may have a similar effect.  For this reason, we perform independent fitting of each spectral line (or doublet) to supplement the results of \rto.

\subsection{MUSE/VLT Spectra}
\label{section:museobs}

We selected all sources from the \rto catalogs that have estimated redshifts $ 2.9 < z < 6.7 $ so that \lya falls within the wavelength range of MUSE (1082 spectra across all 12 clusters) and retrieved the corresponding spectra from the database provided by the authors of \rto.  These spectra were extracted across the entire lensed image (segmentation mask) of each source.  In the case of sources with HST continuum detections, the spectral extraction from the MUSE cubes is weighted by HST continuum surface brightness. In the case of sources detected only via emission lines in the MUSE cubes, the segmentation map of the brightest emission line is used.  \rto also perform subtraction of the local background for each of these spectra, which helps to reduce contamination from foreground galaxies and intracluster light.  Further details of spectral extraction can be found in \rto.

\subsection{Initial Spectral Fitting}
\label{section:fitting}

\begin{figure*}[ht!]
	\centering
	\includegraphics[width=\textwidth]{"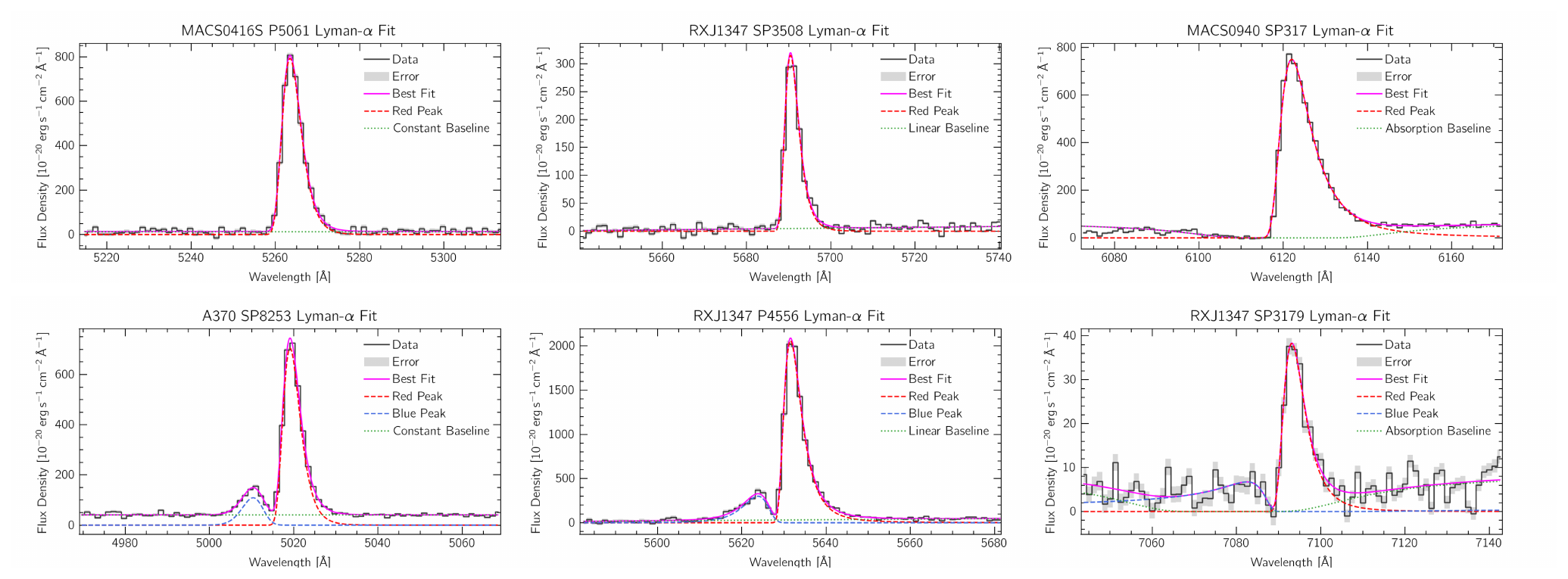"}
	\caption{Example \lya profiles with best fit models. (a) Single-peaked \lya with a constant baseline, (b) single-peaked \lya with a linear baseline, (c) single-peaked \lya with a damped absorption baseline, (d) double-peaked \lya with a constant baseline, (e) double-peaked \lya with a linear baseline, (f) double-peaked \lya with a damped absorption baseline.}
	\label{fig:example_profiles}
\end{figure*}

	For each spectrum, we fitted analytic profiles to all lines listed in the \rto catalogs using \texttt{scipy.curve\_fit} \citep{scipy}.  \lya profiles were modelled with asymmetric Gaussians
	
	\begin{equation}\label{equation:asymmetric_Gaussian}
		f(\lambda) = A \exp \left( - \frac{(\lambda - \lambda_c)^2}{2\sigma_{\text{asym}}^2} \right)
	\end{equation}

	where $A$ and $\lambda_c$ are the amplitude and peak wavelength respectively, and
	
	\begin{equation}\label{equation:sig_asym}
		\sigma_{\text{asym}} \equiv a(\lambda - \lambda_c) + d 
	\end{equation}
	
	with $a$ the asymmetry parameter and $d$ the characteristic dispersion.  For each spectrum, we tried both single- and double-peaked profiles, as well as three different baseline models: constant, linear, or damped-\lya absorption (DLA).  These baselines allowed us to better take into account \lya absorption from neutral gas in the CGM and/or IGM attenuation bluewards of the rest-frame \lya wavelength.  We selected the best \lya model via the Bayesian Information Criterion (BIC).  Full details of the \lya fitting procedure are given in Appendix~\ref{lya_fitting_appendix}.  Examples of each profile type are shown in Figure~\ref{fig:example_profiles}, and a full breakdown of the frequency of each model type is given in Table~\ref{table:lya_models}.  Parameter uncertainties for \lya were derived via bootstrap resampling (1000 iterations).

	\begin{deluxetable}{cccc}
	\tablecaption{Breakdown of the different model types used to fit the \lya profiles in our sample in terms of baseline types and number of fitted peaks.}
	\tablehead{
		\colhead{} & \colhead{Single-peaked} & \colhead{Double-peaked} & \colhead{Total}
	}	 
	\startdata
	Constant & 258 & 75 & 333 \\
	Linear   & 107 & 44 & 151 \\
	DLA      & 56  & 33 & 89  \\
	\hline
	Total    & 421 & 152 & 573
	\enddata
	\end{deluxetable}
	\label{table:lya_models}

	Other emission and absorption lines were each fitted with a Gaussian plus a linear local continuum.  Doublets were fitted simultaneously with shared kinematics, and fits to key diagnostic lines (\civ, \siii, \oiii, \siiv, \ciii, \heii) were forced even when absent from the \rto catalog.  Uncertainties were obtained from the covariance matrix; any $\geq 3\sigma$ detections are treated as tentative and followed up with bootstrap fitting as needed.  Full details of the non-\lya fitting procedure are given in Appendix~\ref{app:other_fitting}.

\subsection{Quality Control}
\label{section:qc}

Following fitting, we performed two rounds of quality control on the fitted line profiles.  First, we identified and removed \lya spectra highly contaminated by systematics and/or artifacts by ranking outliers according to their Local Outlier Factor \citep[LOF;][]{lof_paper}, which decides outlier scores for each point based on the relative density of points in its immediate neighborhood in $N$-dimensional space. We manually chose an LOF cutoff above which profiles were judged unreliable, removing 30 spectra and leaving 1052.  Second, as the target sources lie behind galaxy clusters, we assessed whether spectral features from foreground objects could be imprinted on the target spectra and mistaken for intrinsic features.  For each source we identified the most likely contaminating foreground object, modelled its contribution at the target position, and flagged any fitted line that could plausibly be attributed to the contaminant.  In total, 59 putative line detections across 30 spectra were flagged and excluded from further analysis.  Full details of both quality control procedures are given in Appendix~\ref{app:qc}.

\subsection{Identifying and Aggregating Lensed Counterparts}
\label{section:counterparts}

To maximize the signal-to-noise of our spectra and prevent repeated entries in our catalog, we aggregated spectra from multiply-lensed images of the same source. We used the lensed counterpart assignments given in the \rto catalogs, which were identified via lens modelling combined with manual spectral inspection, to define groups of counter-images. Within each group, we discarded any member whose spectrum showed likely contamination in any of the following critical lines: \nv, \siii, \siiv, \civ, \heii, \oiii, \siiii, or \ciii\ (see Section~\ref{section:qc} for details of how contamination is identified). Any contamination flags in other spectral lines were carried forward to stacked spectrum.  The remaining members were combined via an inverse-variance weighted mean, where each spectrum was weighted by the reciprocal of its median variance. Uncertainties on the co-added spectrum were propagated in quadrature using the same weights. The magnification assigned to each co-added source was taken as the corresponding weighted mean of the individual counter-image magnifications. All spectral lines were then re-fitted using the procedure described in Section~\ref{section:fitting}, with the best-fit parameters from the initial round of fitting used as starting guesses. After consolidating multiply-lensed images in this way, our catalog contains $654$ spectra.

\subsection{Final Fitting and Flagging}
\label{section:final_fit}

\begin{figure*}[ht!]
	\centering
	\includegraphics[width=0.95\textwidth]{"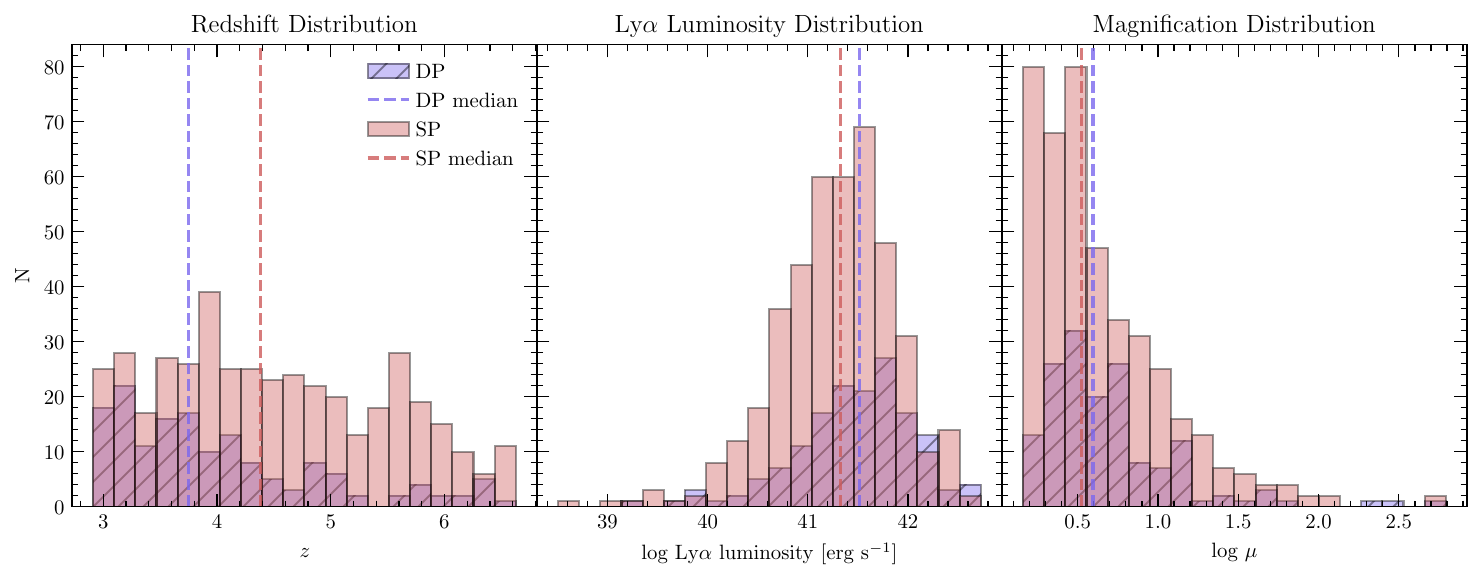"}
	\caption{\textit{Left:} redshifts of the peaks of the fitted \lya profiles (in cases where the \lya is double-peaked, we use the peak wavelength of the redshifted peak).  \textit{Middle:} \lya luminosities (summing both peaks in double-peaked cases) adjusted for lensing magnification based on the lens models of \rto.  \textit{Right:} Magnifications from the \rto lens models.}
	\label{fig:basic_stats}
\end{figure*}

	We re-fitted any candidate metal emission or absorption line with $ |\mathrm{SNR}| > 3 $ in the preliminary round of fitting, using bootstrap resampling (1000 iterations) to robustly estimate uncertainties in fitted parameters.  To help take into account systematics (which can arise from, e.g., residual sky lines or contaminating features from foreground cluster members), we included the effects of autocorrelation in our bootstrap resampling approach (see Appendix~\ref{app:autocorrelation}).  In brief, we used the autocorrelation function of the fit residuals to estimate a correlation length $ \tau $, drew noise realizations where noise of a given point includes a contribution from the previous point according to $ \phi = e^{-1/\tau} $, and then added this correlated noise to the observed spectrum. We took the parameter uncertainties to be the $7\sigma$-clipped standard deviation of the distributions obtained from fitting to each of these 1000 noise realizations.

	We subjected detected lines to automated quality-control flagging to identify potentially spurious detections arising from systematics such as residual sky lines, unresolved noise spikes, or poor Gaussian fits.  Up to six flag types were applied, testing for: (i) proximity to known sky lines, (ii) unresolved line width, (iii) absorption lines dominated by negative flux density, (iv) line profiles dominated by a single spectral channel, (v) unexplained features in the spectrum of comparable strength to the putative lines, and (vi) pre-existing contamination flagged in the initial quality control step in Section~\ref{section:qc}.  When both components of a doublet were independently detected at $|\mathrm{SNR}| \geq 3$, we cleared all flags, since a coincident doublet detection constitutes strong evidence for a genuine feature.  Full details of each flagging criterion are given in Appendix~\ref{app:flagging}.

\subsection{Final Sample}
\label{section:finalselection}

In Figure \ref{fig:basic_stats}, we show distributions in redshift, \lya luminosity, and lensing magnifications (from the lens models of \rto) for our final sample of 573 \lya-emitting sources (counting only those with \lya SNR $ > $ 5), separated into sources fitted with single-peaked and double-peaked \lya profiles.  Though the number of spectra is much reduced by the selection and stacking processes outlined above, almost the full redshift range of is still included in the sample, with the lowest redshift being $ 2.91 $, and the highest being $ 6.63 $. The catalog also contains sources with both high and low redshift confidence as identified by \rto : 107 sources with $ zconf = 1 $, 199 with $ zconf = 2 $, and 267 with the highest confidence of $ zconf = 3 $.  We did not exclude low-confidence sources \textit{a priori}, as we consider our fitting and flagging procedures robust enough to prevent interloper sources from contaminating our results.

\section{Results and Analysis}
\label{section:results}

\subsection{Outflow Signatures}
\label{section:outflowsigs}

In this section, we establish evidence for widespread outflows among our sample of lensed \lya-emitting galaxies both indirectly, via \lya line profiles, and directly, via blueshifted interstellar absorption features, in both individual and stacked spectra.

\subsubsection{Systemic Redshifts}
\label{section:sysvels}

\begin{figure}[ht!]
	\includegraphics[width=0.47\textwidth]{"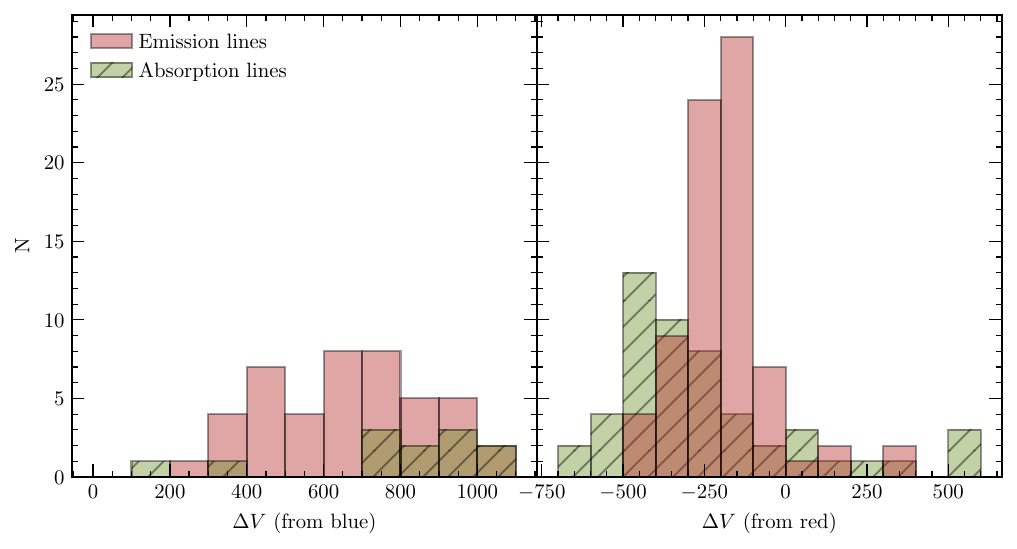"}
	\caption{Velocity offsets of emission and absorption lines relative to the blue \lya peak, where present, (left) and red \lya peak (right).  In spectra where multiple emission or absorption lines are detected, the velocity offsets are calculated as the SNR-weighted average of all such lines.}
	\label{fig:absvsemvel}
\end{figure}

When searching for spectral signatures of outflows (both in \lya profiles and down-the-barrel absorption lines) it is imperative to determine the systemic redshift of the host galaxies.  This can be done most readily using optically-thin emission lines such as \ciii and \heii, which are unaffected by RT effects, unlike resonant lines such as \lya and \civ.   For sources where our initial fitting routine detected two or more optically-thin emission lines\footnote{any combination of \heii, \oiii, \siiii, and/or \ciii} at SNR $ > 5\sigma $, we stacked together all such (un-flagged) lines in velocity space, taking the redshift of the \lya line peak as a temporary proxy for the systemic redshift (or the peak of the more redshifted component in double-peaked cases).  We then fitted the stacked emission spectra with Gaussian profiles to find the velocity offset relative to the \lya peak, this time generating more robust uncertainties via bootstrapping with 1000 noise realizations, and subjecting the fitted lines to the same quality control flagging steps as outlined in Appendix \ref{app:flagging} (with the exception of the flags for sky lines and unresolved lines, which only apply to fits in wavelength space).  We also searched for stacked optically-thin emission from sources where our initial fitting found fewer than two individual emission lines at SNR $ > 5\sigma $.  To do this, we stacked the \ion{O}{3}]$ \lambda 1666 $, \ciii, and \heii lines -- the most commonly found optically-thin emission lines in our initial fitting -- and fitted a Gaussian, again with bootstrapping to generate uncertainties and flagging.  Using these methods, we were able to measure the systemic redshift in 81 sources (out of 573), allowing us to explore key outflow diagnostics from \lya emission and metal absorption lines.

\subsubsection{\lya Profiles}
\label{section:lyaprofiles}

We find a diverse range of \lya profiles, ranging from extreme single-peaks to double-peaks that approach comparable height.  Henceforth, we refer to sources fitted with single- and double-peaked \lya profiles as SP and DP sources respectively -- though we caution that SNR and attenuation by neutral gas in the IGM play a role in deciding which of these categories a source belongs to.  Out of the 573 spectra, 421 (73\%) are SP and 152 (27\%) are DP (i.e. we find two peaks each with significance $ > 3\sigma $), including 36 sources with $z > 4$ shown in Appendix~\ref{app:highz_dp}.  Throughout this paper, we refer to the more redshifted \lya emission component as the ``red peak'', and (where present) the less redshifted component as the ``blue peak''.  SP and DP profiles form a continuum in this sense: the red peak is the same physical entity in both cases, with the blue peak in DP sources being an additional secondary feature that arises when outflow conditions permit detectable blueshifted \lya photons to escape.

The \lya profile morphologies are, in the overwhelming majority of cases, consistent with outflows.  Considering first the DP sources, in all but four cases the red peak integrated flux is greater than that of the blue peak.  In all four cases where the blue peak flux exceeds the red, the flux ratio between the peaks is within $ \sim 2\sigma $ of 1, i.e. there is no strong evidence for genuinely blue-dominated profiles.  More generally, the red peak is positively asymmetric (red-skewed) or consistent with positive asymmetry at the $ 3\sigma $ level in 97\% of sources regardless of SP/DP classification -- a hallmark of \lya profiles shaped by outflowing gas \citep[e.g.][]{verhamme06}.

In cases where the systemic redshift can be measured, the evidence for outflows becomes clearer.  The left and right panels of Figure \ref{fig:absvsemvel} show the velocity offsets of the optically-thin emission lines relative to the blue (where present) and red \lya peaks, respectively.  In the vast majority of sources with measured systemic redshifts (74 out of 81), the more redshifted \lya component\footnote{In SP sources there is only one component, so this simply means the \lya line itself.} is redshifted relative to the systemic.  Among the 46 DP sources with measured systemic redshifts, the two \lya peaks straddle the systemic redshift in all but one case (where both peaks are blueshifted relative to the systemic).  In the 35 SP sources with systemic redshifts, in all but six cases the \lya line is redshifted relative to the systemic.  In total, seven sources show ``red" \lya peaks that are blueshifted relative to the systemic. These could represent genuine cases of inflowing gas, or alternatively contamination from unresolved merging or interacting galaxy pairs at slightly differing redshifts.  A full analysis of these exceptional cases is deferred to a future study.

Taken together, these trends agree with previous work by \cite{kulas12} that measured the systemic redshifts of \lya emitters at $ z = 2 $--$ 3 $, and show broad qualitative agreement with expanding shell RT models, suggesting widespread outflows from these sources.

\subsubsection{Blueshifted Metal Absorption Lines}
\label{section:abslines}

\begin{figure*}[th!]
	\includegraphics[width=0.95\textwidth]{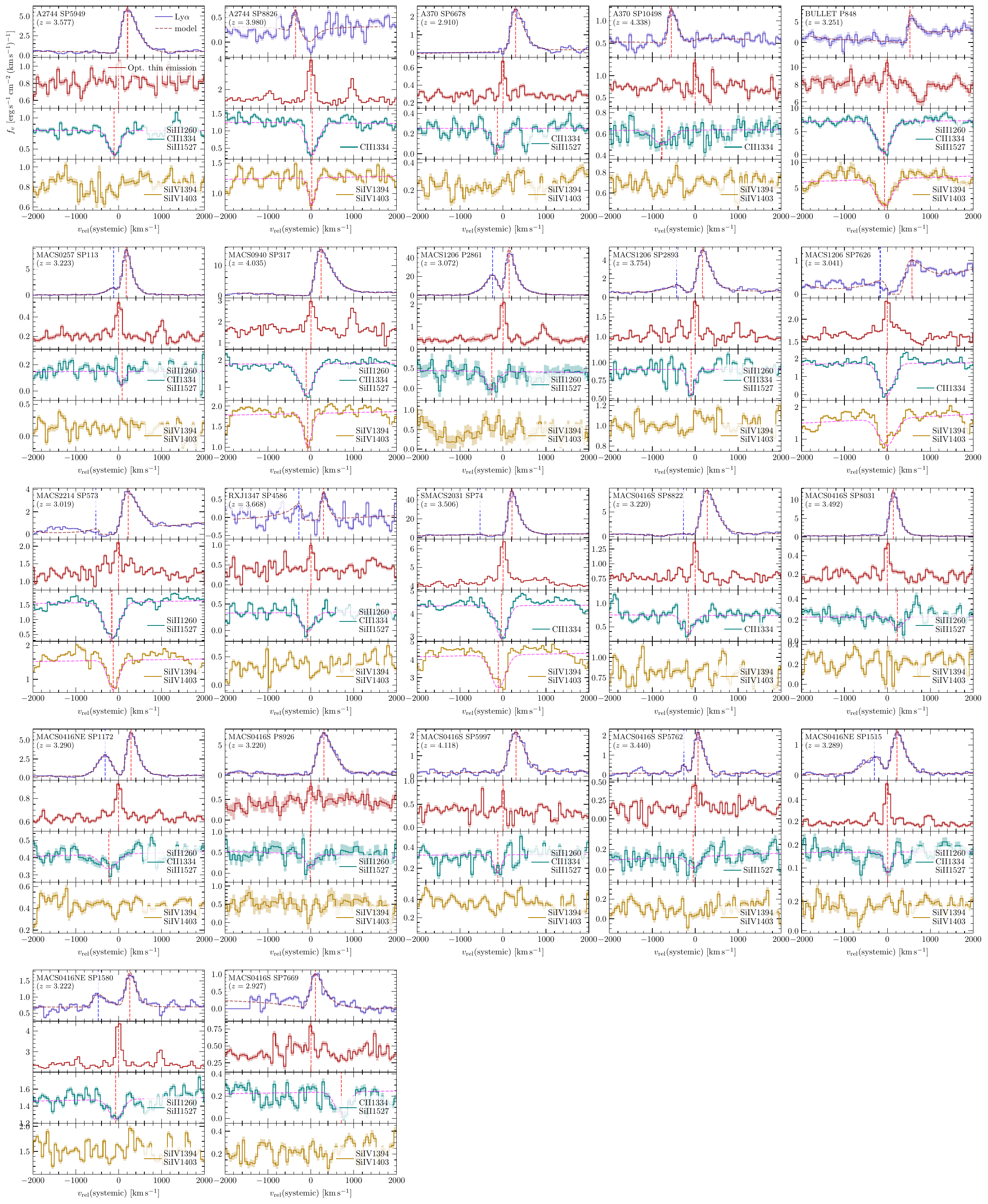}
	\caption{Comparison in velocity space of \lya profiles, stacked optically-thin emission lines, stacked low-ionization absorption, and stacked high-ionization absorption profiles. The zero point of the velocity axis is centered on the systemic velocity, and the central velocities of the Gaussians fitted to the absorption profiles (where detected) are shown with dashed vertical red lines.}
	\label{fig:absorbers0}
\end{figure*}

We now analyze the metal absorption lines found in our sample.  As we will show, we find kinematics indicative of widespread outflows: (i) overall, absorption lines tend to show significantly greater velocity offsets relative to \lya than optically thin emission lines; (ii) in cases where the systemic redshift can be measured, absorption lines are blueshifted (or consistent with blueshifted) in the majority cases (18 out of 22).  Furthermore, we find evidence for clumpy outflow geometry based on absorption line ratios of both high- and low-ionization ionic species.

To search for signs of outflows, we made use of the metal absorption lines \siii, \cii, and \siiv.  These were by far the most common lines found by our initial fitting routine, and have been used to probe the kinematics of outflowing gas in numerous studies of SFGs (\citealt{shapley03,steidel10,jones12,berry12}). \ion{Si}{2} and \ion{C}{2} are associated with predominantly neutral gas due to their lower ionization potentials ($ < \SI{13.6}{eV} $), while \ion{Si}{4} traces the ionized phase with temperatures $  \SI{10^4}{K} \lesssim T \lesssim \SI{10^5}{K} $ (see, e.g., \citealt{sutherland93}).  To characterize the overall outflow kinematics in each source with maximum SNR, we stacked together lines belonging to the same group (low- or high-ionization) and fitted Gaussian profiles with bootstrapped uncertainties accounting for autocorrelation in the spectra and quality control flagging as in Section \ref{section:final_fit}.  We performed this procedure on three combinations of different low-ionization lines: \cii alone, \siii stacked together with \siiih\footnote{It is not explained why \rto did not include the \siiih line in their spectral templates, but it is reported among other low-ionization lines in the spectra of \lya-emitting galaxies by other authors, e.g., \cite{berry12}.}, and all low-ionization lines: \cii, \siii, and \siiih.  We then chose the combination with the highest SNR fit (\ion{C}{2} only, \ion{Si}{2} only, or both species combined) to represent the low-ionization interstellar absorption.  Similarly, we stacked both lines of the high-ionization \siiv doublet and fitted the resulting spectrum.  We discarded putative fits involving negative flux density (which are likely due to residual sky lines) as well as those where no continuum source was detectable in an image made by averaging all the spectral channels between $ 1220 $ and $\SI{1380}{\AA}$ in rest frame (genuine absorption being highly unlikely with such weak continuum). In total, we find $ 39 $ sources with statistically significant ($ > 3\sigma $) absorption at redshifts ranging from $ z = 2.9 $ to $ z = 5.7 $, of which  and $ 11 $ have high-ionization absorption. In nine sources, both low- and high-ionization absorption are detected independently.

The kinematics of these absorption profiles are consistent with outflows in most cases: all but seven are blueshifted relative to the red \lya peak as shown in Figure \ref{fig:absvsemvel} (green histogram).  Furthermore, when compared with optically-thin emission lines, absorption lines have a clear tendency to have greater velocity offsets relative to the \lya peak, as may be expected if emission lines trace the systemic velocity while absorption lines trace outflowing gas.  $ 22 $ of the sources in which we detect absorption also have systemic redshifts measured via stacked optically thin nebular emission lines, allowing us to explicitly measure absorption kinematics relative to the host galaxy. Of these 22 sources, all have redshift confidence from \rto of $zconf > 1$, except for MACS0416S SP7669, which we consider to be a strong case for genuine \lya emission due to its highly asymmetrical \lya profile with tentative blue peak.

\begin{figure}[ht!]
	\includegraphics[width=0.48\textwidth]{"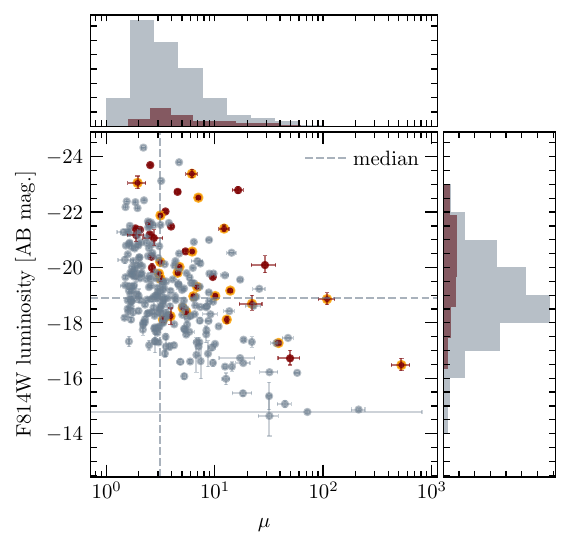"}
	\caption{Luminosity in \emph{HST/ACS} F814W band vs lensing magnification for sources with (red) and without (grey) detectable interstellar absorption.  Sources where the systemic velocity can also be measured are highlighted orange.  Median values for the non-absorbers are shown with dashed lines.}
	\label{fig:mumagcomp}
\end{figure}

In Figure \ref{fig:absorbers0}, we compare the \lya, stacked optically-thin emission, and low- and high-ionization absorption lines in each of the 22 sources in which absorption was found and the systemic velocity could also be measured.  The centroid of the fitted absorption lines are blueshifted relative to the systemic velocity in all but 5 cases, one of which is consistent with blueshift within measurement uncertainties (MACS0416NE SP1515).  To understand whether these objects are representative of our sample of \lya-emitting galaxies as a whole, we compare sources with detectable absorption lines against the rest of the sample in terms of continuum luminosity in the \emph{HST/ACS} F814W filter and estimated lensing magnification in Figure \ref{fig:mumagcomp}.  As may be expected, some of the sources with detectable absorption are unusually luminous in the continuum ($ \lesssim \SI{-22}{AB\,mag.} $), but many are close to or even below the sample median ($ \SI{-19}{AB\,mag.} $).  Furthermore, most of the sources with detectable absorption have magnifications above the median, highlighting the advantage conferred by gravitational lensing.  These results, along with the observed characteristics of the \lya profiles described in Section \ref{section:lyaprofiles}, suggest that typical \lya-emitting galaxies in the early universe host outflows.

\begin{figure}[ht!]
	\includegraphics[width=0.48\textwidth]{"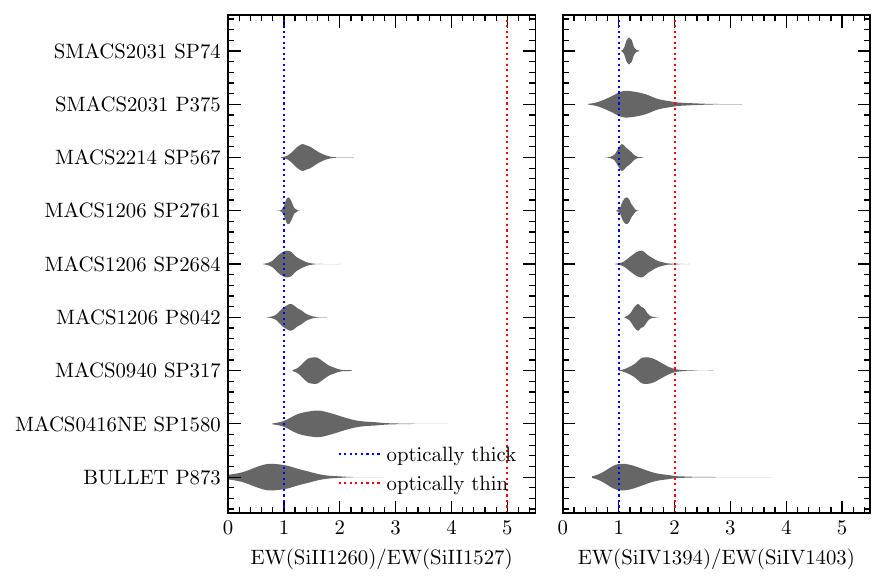"}
	\caption{EW ratios for the low-ionization \ion{Si}{2} absorption doublet (left) and high-ionization \ion{Si}{4} absorption doublet (right).  Distributions are generated by Monte Carlo bootstrapping of the spectral uncertainties when fitting the absorption profiles.  Absorption in optically thick clumps leads to an EW ratio of 1 as indicated by the blue dashed line, while the ratios expected for optically thin absorption are shown with red dashed lines.}
	\label{fig:clumpiness}
\end{figure}

The sources with the strongest absorption signatures allow us to probe the structure of the outflows using ratios between absorption lines of the same ionic species.  Line ratios corresponding to the ratio of the oscillator strengths of the respective lines indicate that the absorption is unsaturated (i.e. the gas is optically thin); on the other hand, line ratios approaching unity indicate that the absorption is saturated (i.e. the gas is optically thick).  Line ratios approaching unity but with non-zero flux density at the line cores indicate saturated absorption in gas with limited covering fraction -- i.e. a clumpy medium.  Using this method, \cite{berry12} find evidence that high-ionization absorption lines in LBGs are optically thin, based on equivalent width (EW) ratios of the \siiv doublet, $\text{EW}(\text{SiIV}1394) / \text{EW}(\text{SiIV}1403) \sim 2.0 $, which is approximately the same as the ratio of oscillator strengths.  This contrasts with the low-ionization lines which are generally found to be consistent with absorption in optically-thick clumps \citep{shapley03, steidel10, berry12, jones12} as measured via the \ion{Si}{2} absorption doublet, where for optically thin absorption the ratio is $\text{EW}(\text{SiII}1260) / \text{EW}(\text{SiII}1527) \sim 5.0 $.  We perform the same measurement by re-fitting the strongest absorption lines.  To get a more complete picture of the uncertainties in absorption line EW ratios, we again use bootstrapping, taking the ratio of EWs for each of the $ 1000 $ noise realizations.

We show the resulting distributions in EW ratios of low- and high-ionization absorption lines in Figure \ref{fig:clumpiness}.  In agreement with past studies of the \ion{Si}{2}$\lambda\lambda1260,1527$ doublet, we find EW ratios $ \sim 1 $, suggestive of optically thick clumps.  On the other hand, in the \siiv doublet we find mixed results, with four of eight sources being inconsistent with the optically thin limit, and all sources preferring an intermediate line ratio between the optically thick and thin limits.  These results are consistent with outflows containing clumps of gas with a range of temperatures and densities.  \ion{Si}{2} can only survive in dense, self-shielded clumps, and hence is optically thick, while \ion{Si}{4} may exist in hotter, more tenuous clumps of lower optical depth.

\subsection{Source vs Outflow Properties}
\label{section:source_vs_outflow}

\subsubsection{\lya Profiles vs Outflow Velocity}
\label{subsection:lya_vs_vout}

\begin{figure*}[ht!]
	\includegraphics[width=\textwidth]{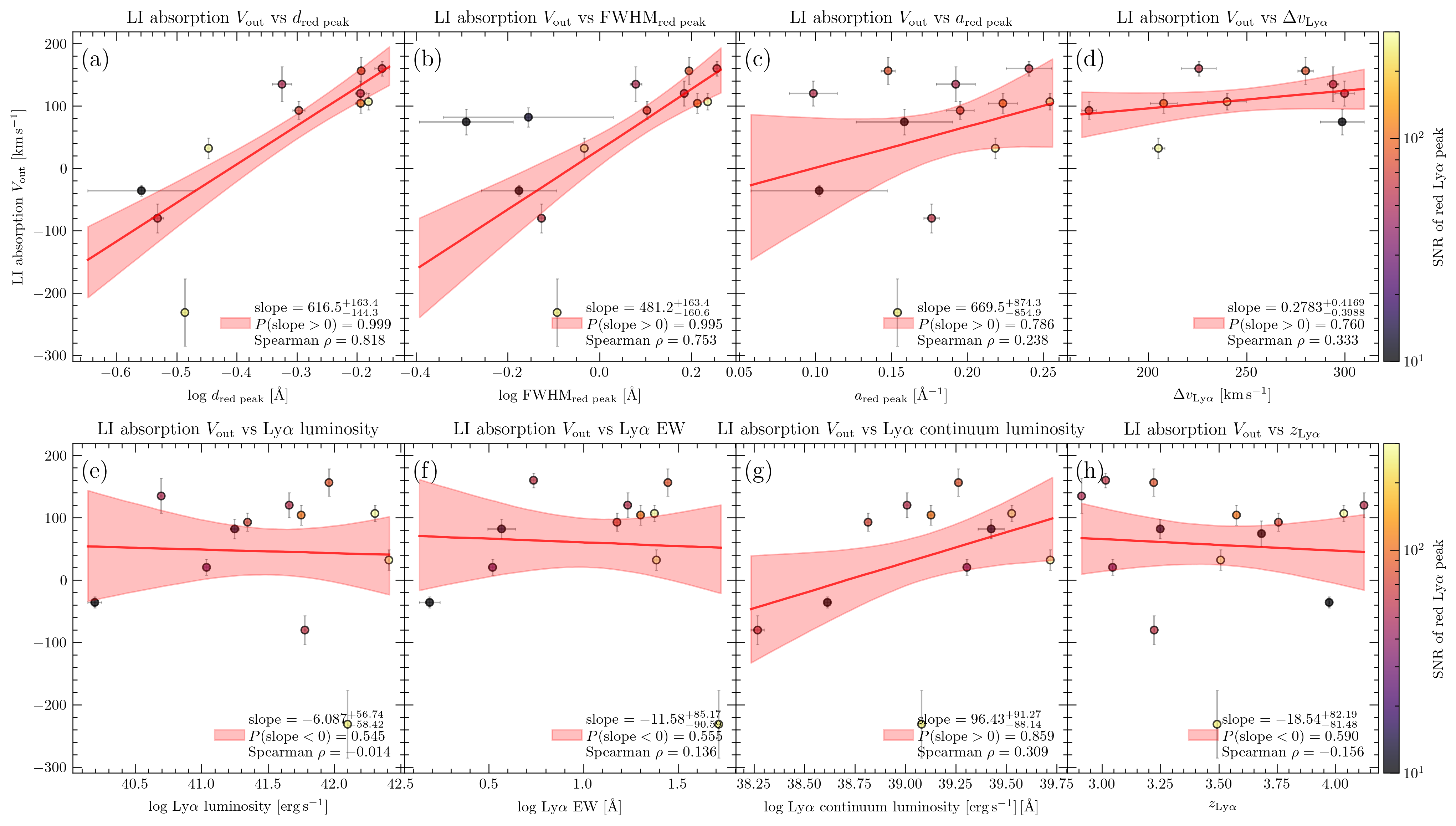}
	\caption{Outflow velocity $ V_{\mathrm{out}} $ measured from low-ionization metal absorption lines versus \lya profile parameters: (a) intrinsic line dispersion $ \sigma_{\mathrm{Ly\alpha}} $, (b) FWHM, (c) velocity offset of the \lya red peak from systemic $ \Delta V_{\mathrm{Ly\alpha}} $, (d) blue--red peak separation $ \Delta\lambda_{\mathrm{BR}} $ (double-peaked sources only), (e) \lya luminosity, (f) \lya EW, (g) luminosity of the continuum adjacent to the \lya line, and (h) redshift of the red \lya peak.  The \texttt{LinMix} regression is shown with its 68\% posterior credible interval (shaded band).  Data points are colored by integrated SNR of the red \lya peak.}
	\label{fig:vout_dispr_corr}
\end{figure*}

We now investigate whether the shape of the \lya line encodes information about outflow kinematics by comparing \lya profiles with metal absorption line profiles.  We do this by searching for correlations between outflow velocity as defined by the centroid of the fitted Gaussian absorption profile, $ V_{\mathrm{out}} $, and the following eight observed \lya profile parameters: intrinsic red peak line dispersion $ d_{\mathrm{Ly\alpha}} $, red peak FWHM, red peak asymmetry $ a_{\mathrm{Ly\alpha}} $, offset of the \lya peak relative to the systemic velocity $ \Delta v_{\mathrm{Ly\alpha}} $,  \lya luminosity, \lya equivalent width, luminosity of the continuum adjacent to the \lya line, and \lya peak redshift\footnote{All parameters are converted to rest-frame values, corrected for lensing magnification, and corrected for instrumental dispersion where necessary.}.  We use only those sources in which $ V_{\mathrm{out}} $ could be measured from low-ionization metal absorption lines relative to the systemic redshift (Section~\ref{section:abslines}) with high statistical significance (absorption detected at $ \geq 5 \sigma$).  We perform Bayesian linear regression using the \texttt{LinMix} package \citep{linmix}, which accounts for measurement uncertainties in both variables and models the intrinsic scatter with a mixture of Gaussians\footnote{We adopted $ K = 2 $ Gaussian components for the mixture model prior on the distribution of the independent variable, following \cite{linmix}}.

To lessen the influence of extreme outliers in our plots -- which can both induce spurious correlations and destroy genuine ones -- we prune the data by (i) removing points that deviate from the sample mean in the $x$ or $y$ axis by more than $ 3 $ times the sample standard deviation along the same axis, and (ii) removing points whose normalized mean Euclidean distance from all other points satisfies

\begin{equation}
\bar d_i > \operatorname{median}(\bar d) + 3\,\operatorname{MAD}(\bar d)
\end{equation}

where

\begin{equation}
\bar d_i = \frac{1}{N-1}\sum_{j \ne i}\sqrt{\left(\frac{x_i-x_j}{\sigma_x}\right)^2 + \left(\frac{y_i-y_j}{\sigma_y}\right)^2}
\end{equation}

We also remove data points whose $x$ or $y$ error bar exceeds the standard deviation of the data points along the corresponding axis, as such points have little constraining power.

The scatter plots showing $ V_{\mathrm{out}} $ versus along with lines of best fit are shown in Figure \ref{fig:vout_dispr_corr}. We evaluate each correlation using the posterior probability that the regression slope has the reported sign, i.e.\ $P(\mathrm{slope} > 0)$ for positive correlations and $P(\mathrm{slope} < 0)$ for negative correlations.  We additionally quote the Spearman rank correlation coefficient $ \rho $ as a model-independent, scale-free characterization of association strength.  We find strong posterior support for a positive correlation between $ V_{\mathrm{out}} $ and $ d_{\mathrm{Ly\alpha}} $ ($P(\mathrm{slope} > 0) = 0.999 $; Spearman $ \rho = 0.818 $), as illustrated in panel~(a) of Figure~\ref{fig:vout_dispr_corr}.  Sources with broader \lya line profiles therefore harbour faster outflows as traced by low-ionization absorption lines.  This correlation stands in qualitative agreement with expectations of expanding shell RT models, in which faster outflowing shells scatter \lya photons to larger velocity offsets, thereby broadening the emergent profile \citep{verhamme06,orsi12,li22}.  A consistent result is found between $ V_{\mathrm{out}} $ and \lya FWHM ($P(\mathrm{slope} > 0) = 0.995$; Spearman $ \rho = 0.753 $; panel~(b)), though this is unsurprising as the FWHM is closely linked to $ d_{\mathrm{Ly\alpha}} $ via equations \ref{equation:asymmetric_Gaussian} and \ref{equation:sig_asym}.  We also find weak plausible correlations between $ V_{\mathrm{out}} $ and $ a_{\text{red peak}} $ as well as $ \Delta V_{\mathrm{Ly\alpha}} $ ($P(\mathrm{slope} > 0) \sim 0.8$ in both cases; panels~(c) and (d)). While these correlations have limited posterior support, they also agree qualitatively with expanding shell RT models.

Together, these results suggest that \lya profile width encodes quantitative information about outflow kinematics.  In particular, they suggest that faster outflows lead to \lya profiles that are more ``stretched out" along the wavelength/velocity axis, leading to greater line widths and offsets from systemic velocity.  Moreover, these trends all qualitatively agree with simple expanding shell/wind RT simulations, representing a much needed empirical check of those models.

Among the remaining \lya parameters (\lya luminosity, \lya EW, \lya continuum luminosity, and $ z_{\mathrm{Ly\alpha}} $), only the \lya continuum luminosity shows a weak correlation, though this is entirely driven by the two data points with the lowest continuum luminosities.  Excluding these points, the correlation reverses sign entirely, and we therefore do not consider this to be genuine.  It is notable that neither \lya luminosity nor \lya EW show any relation with outflow velocity -- we discuss this further in Section \ref{subsection:lya_vs_em}.

\subsubsection{\lya Profiles vs Stellar/ISM Properties}
\label{subsection:lya_vs_em}

With a demonstration that \lya profile morphology encodes genuine outflow kinematic information (Section~\ref{subsection:lya_vs_vout}), we now investigate what stellar and/or ISM properties drive the differences in outflow velocity inferred from \lya profiles.

We searched for correlations between the EW of nebular emission lines and the same \lya parameters that we established as having a relation with outflow velocity in Section~\ref{subsection:lya_vs_vout} using the exact same methodology as before.  We also looked at an additional \lya parameter, namely the rest-frame separation between blue and red peaks in double-peaked sources, $ \Delta v_{\mathrm{blue-red}} $ (we did not include this parameter when searching for correlations with outflow velocity due to there being an insufficient number of double-peaked sources, just six, with detectable absorption).  Here, we used only those emission lines where both the line itself and adjacent continuum were detected with high statistical significance (SNR $ > 5$).  In all cases where lines formed part of a doublet, we summed the line together with its doublet partner in deriving the EW, and applied redshift corrections to obtain rest frame values.

\begin{figure*}[ht!]
	\includegraphics[width=\textwidth]{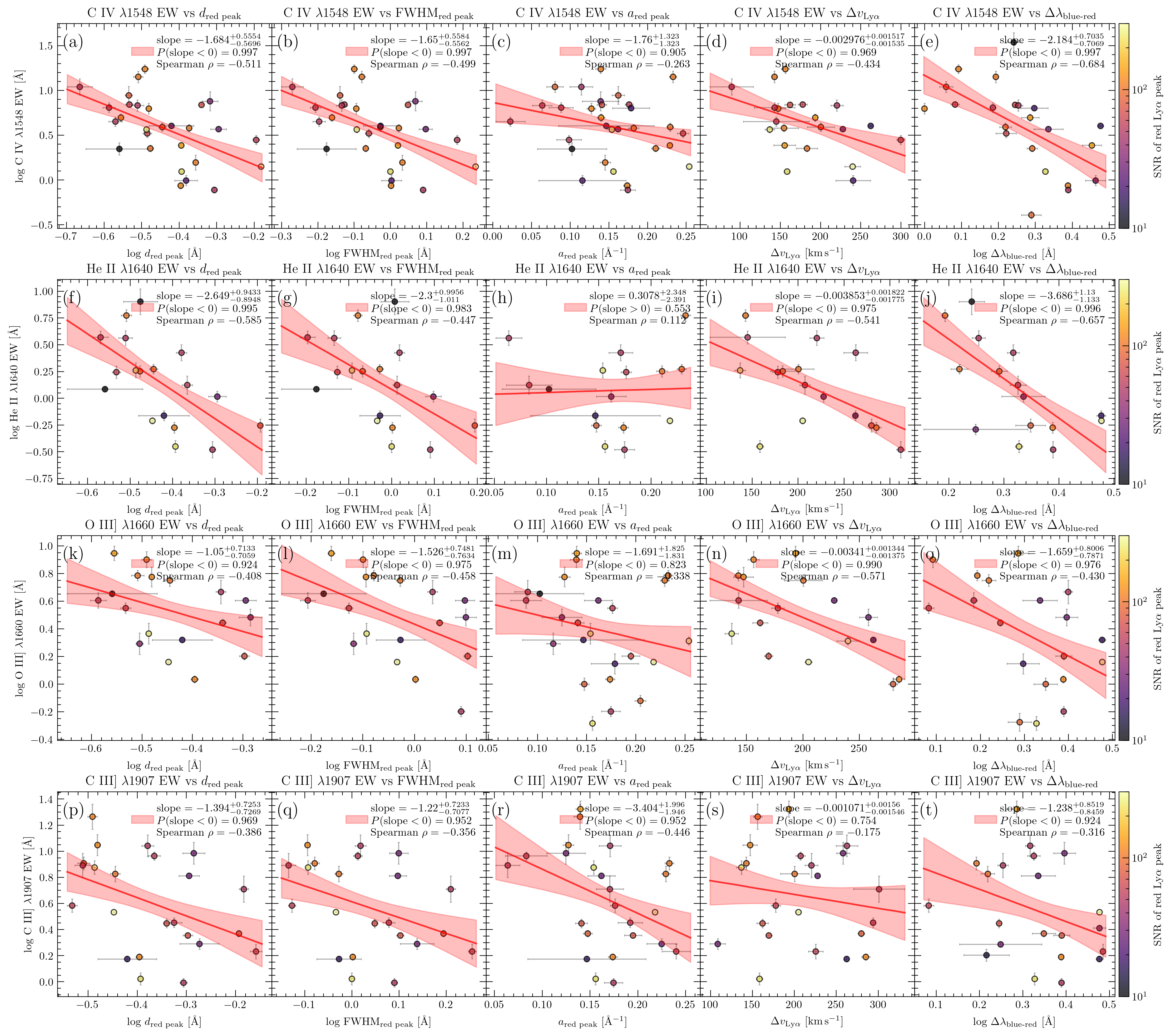}
	\caption{Metal emission line EWs versus various \lya dispersion measures as well as red peak asymmetry. Linear regression results are shown in red as in Figure~\ref{fig:vout_dispr_corr}. Note that the \ion{C}{4}, \ion{O}{3}], and \ion{C}{3}] lines have all been summed with their doublet partners.}
	\label{fig:dispr_ew_corr}
\end{figure*}

We show the scatter plots along with models produced by \texttt{LinMix} in Figure~\ref{fig:dispr_ew_corr}.  A number of striking trends can be seen, namely:

\begin{enumerate}

	\item  \textit{Emission line EWs are anticorrelated with $ d_{\text{red peak}} $ and \lya FWHM}. \civ (panel~a) and \heii (panel~f) emission show the strongest trends with respect to $ d_{\text{red peak}} $, with $ P(\mathrm{slope} < 0) = 0.997 $ and $ 0.995 $ respectively.  \oiii (panel~k) and \ciii (panel~p) show consistent trends though with greater scatter.  These results indicate that sources with stronger nebular emission relative to the continuum tend to display \emph{narrower} \lya profiles.

	\item Similarly, \textit{emission line EWs are anticorrelated with \lya blue-to-red peak separation $ \Delta v_{\mathrm{blue-red}} $}. Again, \civ (panel~e) and \heii (panel~j) EW are most strongly anticorrelated with $ \Delta v_{\mathrm{blue-red}} $, while \oiii (panel~o) and \ciii (panel~t) show consistent correlations but with greater scatter.

	\item \textit{Emission line EWs are anticorrelated with \lya red peak velocity offset $ \Delta v_{\mathrm{Ly\alpha}} $, albeit more weakly}. \civ (panel~d), \heii (panel~i), and \oiii (panel~n) all show tentative correlations with $ 0.97 \lesssim P(\mathrm{slope} < 0) \lesssim 0.99 $.  \ciii (panel~s) shows considerably more scatter, with $ P(\mathrm{slope} < 0) \sim 0.75 $, though the trend is consistent with the other emission lines.

	\item \textit{Emission line EWs show only weak relationships with \lya asymmetry, $ a_{\text{red peak}} $}.  While \ciii EW shows a tentative anticorrelation with $ a_{\text{red peak}} $ (panel~r), the remaining lines show only very weak anticorrelations or, in the case of \heii (panel~h), no sign of a relation whatsoever.

\end{enumerate}

Taken together, these results demonstrate a clear anticorrelation between the dispersion -- including $ \Delta v_{\mathrm{Ly\alpha}} $ and $ \Delta \lambda_{\mathrm{blue-red}} $ -- of the \lya profile and the EW of nebular emission lines.  Based on the results presented in Section~\ref{subsection:lya_vs_vout} and theoretical expectations from expanding shell models, greater \lya dispersion implies greater outflow velocity.  Therefore, the anticorrelations shown in Figure~\ref{fig:dispr_ew_corr} suggest that \textit{\lya-emitting galaxies with more intense ionization conditions drive slower outflows.} We consider what stellar population/ISM properties underpin this relationship in Section~\ref{subsection:ew_factors}.

We find that the same correlations hold even when excluding sources assigned low redshift confidence by \rto ($zconf = 1$), though some have slightly diminished statistical confidence (e.g. \heii EW vs $d_{\text{red peak}}$ $P(\mathrm{slope} < 0)$ decreases from 0.995 to 0.992). The worst-affected correlations are \civ EW vs $\Delta v_{\mathrm{Ly\alpha}}$, where $P(\mathrm{slope} < 0)$ decreases from 0.969 to 0.846, and \civ EW vs $d_{\text{red peak}}$, where $P(\mathrm{slope} < 0)$ decreases from 0.997 to 0.976. These changes are disproportionately driven by the loss of a single source, MACS0416NE M177, which has the knock-on effect of triggering the removal of another source with $zconf = 3$ due to our outlier removal algorithm. Inspecting the \civ and \lya fits for MACS0416NE M177 reveals it to be a visually-convincing case of a LAE, though other lines such as \ciii and \heii are strongly affected by sky lines, possibly explaining the low redshift confidence assigned by \rto.

\begin{figure*}[ht!]
	\includegraphics[width=\textwidth]{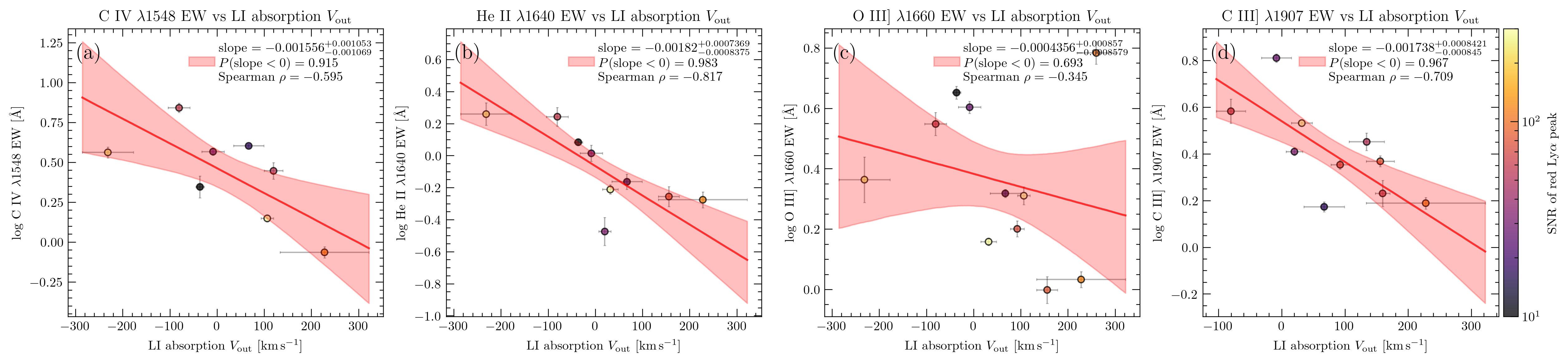}
	\caption{Emission line EWs versus outflow velocity based on low-ionization absorption lines. As before, linear regression results are shown in red. Note that the \ion{C}{4}, \ion{O}{3}], and \ion{C}{3}] lines have all been summed with their doublet partners.}
	\label{fig:ew_vout_corr}
\end{figure*}

We also searched for direct anticorrelation between emission line EWs and $ V_{\mathrm{out}} $.  Though the number of sources with reliable measurements of both $ V_{\mathrm{out}} $ and emission line EWs is small, this represents a useful consistency check.  The results, displayed in Figure~\ref{fig:ew_vout_corr}, show tentative anticorrelation between \heii (panel~b) and \ciii (panel~d) EW and $ V_{\mathrm{out}} $ ($ P(\mathrm{slope} < 0) > 0.95 $) in both cases, and results suggesting the same anticorrelation for \civ (panel~a) and \oiii (panel~c), though with lower confidence.  \oiii is affected by two strong outliers (the points having the lowest and highest values of $V_{\mathrm{out}}$, respectively) that are not caught by the pruning criteria described in Section~\ref{subsection:lya_vs_vout}.  Excluding these outliers, a central sequence of points can clearly be identified, suggesting that the same relationship holds for \oiii as for the other lines.

The only emission line that shows no evidence of a relationship between EW and $V_{\mathrm{out}}$ is \lya itself, as shown in Figure~\ref{fig:vout_dispr_corr}(f). That the trend in \lya is less clear than in other lines is perhaps unsurprising, as \lya EW is modulated by the escape fraction of \lya $f_{\mathrm{esc}}^{\mathrm{Ly\alpha}}$, which in turn depends sensitively on dust and neutral gas in the ISM. Blueshifted \lya emission (where present) is also modulated by any absorption by neutral gas in the IGM, which likely increases the scatter further.

\subsubsection{Factors Determining Emission Line EWs}
\label{subsection:ew_factors}

The results in Sections~\ref{subsection:lya_vs_vout} and \ref{subsection:lya_vs_em} suggest that outflow velocity tends to be lower in systems with greater UV metal emission line EW, which in turn are controlled by stellar population and ISM properties. The fact that \civ and \heii EW generally show the strongest anticorrelations with tracers of outflow velocity points to stellar population age as the crucial factor, as these lines require highly energetic photons of $ \SI{54}{eV} $ and $ \SI{48}{eV} $, respectively, to be produced.  Such energetic photons can only be produced by the youngest and most massive stars, which rapidly evolve away from the main sequence within just a few Myr \citep{feltre16,martins25}. To guard against possible contamination by AGN-powered line emission, we compared emission line ratios against model grids of AGN emission from \cite{feltre16} and \ion{H}{2} region emission from \cite{gutkin16}. We find all cases to be consistent with \ion{H}{2} regions as shown in Appendix~\ref{app:emorigin}, giving us confidence that our metal emission lines indeed trace stellar/ISM properties.

To check whether stellar population age can indeed drive the variation in emission line EW found in our sample, we turned to joint stellar population + ISM models from \cite{gutkin16}, using the \texttt{BEAGLE} Bayesian stellar population fitting package \citep{chevallard16} to fit our observed emission line EWs to the model grids.  We searched for sources where all four diagnostic UV lines (\civ, \heii, \oiii, and \ciii; with doublet components summed where appropriate) were detected at $> 3\sigma$. This yielded 15 sources in total, whose EW values are given in Appendix~\ref{app:ew_table}.  Due to the fact that \civ is associated with a resonance transition, we marked this EW as a \emph{lower} bound on the true value.  We also included another commonly detected line, \siiii, which was found in eight of our 15 sources.  In cases where \siiii was not detected, $3\sigma$ upper bounds were used instead (unless the continuum could not even be detected, in which case we reverted to an arbitrary upper bound of 100 -- far above the range of values found in the data).

\begin{deluxetable}{ccc}
	\tablecaption{Summary \texttt{BEAGLE} configuration, including (i) parameter name, (ii) scale (log or linear), and (iii) uniform prior range.}
	\tablehead{
		\colhead{Parameter} & \colhead{Scale} & \colhead{Prior range}
	}	 
	\startdata
	Age [yr] & log & $ [5.0, 8.0] $ \\
	$ Z / Z_\odot $ & log & $ [-2.2, 1.0] $ \\
	$ \log U $ & log & $ [-4.0, -1.0] $ \\
	$ \xi_d $ & linear & \textbf{fixed} $ 0.1 $ \\
	$ (C/O)/(C/O)_{\odot} $ & linear & $ [0.1, 1.0] $
	\enddata
\end{deluxetable}
\label{table:beagle}

\texttt{BEAGLE} allows the user to fit for star formation history (SFH), stellar and ISM metallicity $ Z $, stellar mass $ M_{\star} $, ionization parameter $ U $, dust depletion factor $ \xi_d $\footnote{$ \xi_d $ controls the degree to which abundance of elements in the ISM is depleted by their absorption into dust grains}, carbon-to-oxygen abundance ratio $ C/O $, and dust attenuation. Our aim was to perform a simple qualitative comparison between our 15 sources to understand which fitted parameters scale with emission line EWs, rather than perform detailed analysis of stellar populations.  For this reason, we chose a constant star formation history, and fixed the dust depletion factor $ \xi_d = 0.3 $.  As we are working with EWs only, stellar mass and dust attenuation cannot be constrained. The priors we use to run \texttt{BEAGLE} are given in Table~\ref{table:beagle}.

\begin{figure*}[ht!]
	\centering
	\includegraphics[width=\textwidth]{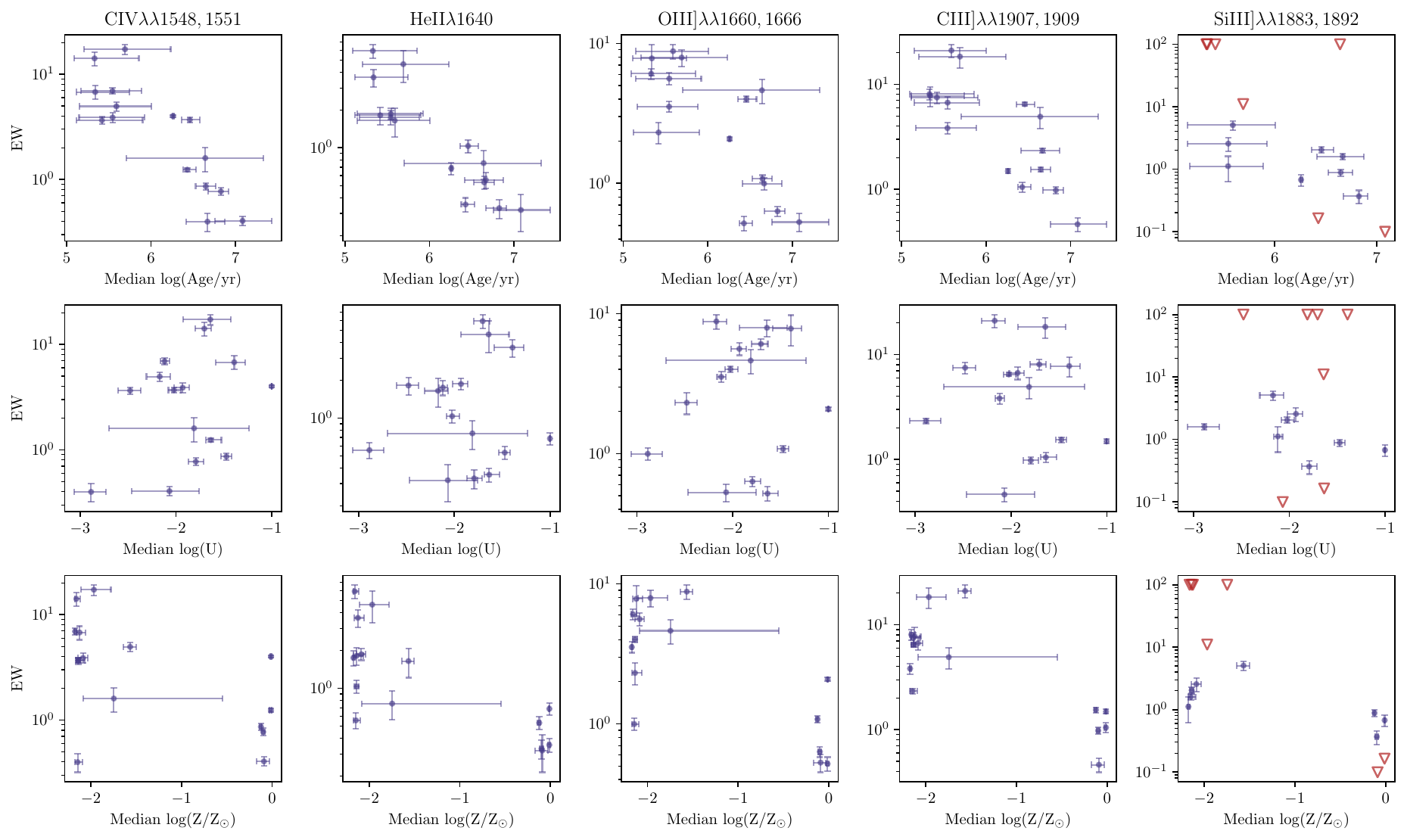}
	\caption{Fitted parameters produced by \texttt{BEAGLE} for the 15 sources with $>3\sigma$ detections in \civ, \heii, \oiii, and \ciii (columns 1 -- 4). \siiii is shown in the final column, and includes some upper limits, indicated by inverted triangles.  Fitted parameter values are obtained from the median of the corresponding posterior distributions. Error bars for EWs are shown at the $1\sigma$ level, while the error bars in the fitted parameters extend to the $16^{\mathrm{th}}$ and $84^{\mathrm{th}}$ percentiles.}
	\label{fig:ew_vs_beagle_params}
\end{figure*}

In Figure~\ref{fig:ew_vs_beagle_params}, we show the emission line EWs used to constrain the models as a function of the fitted parameter values (with the exception of the carbon-to-oxygen abundance, which only affects the relative strengths of the emission lines).  The plots of EW vs age (first row) show a clear trend: sources with larger UV metal line EWs preferentially favor younger stellar ages.  On the other hand, ionization parameter (second row) shows generally positive correlation with emission line EW, though with considerably greater scatter.  Finally, metallicity (third row) shows a highly bimodal distribution, with low- and high-EW sources generally favoring high and low metallicity solutions, respectively.  We do not consider age-metallicity degeneracy to be a major concern here, as the sources with older stellar populations are also those with the highest metallicity.  Given the highly simplistic star formation history we have employed we emphasize that these results are intended to inform a qualitative picture only, rather than a precise determination of galaxy properties. With this in mind, our model results are consistent with the assertion that stellar age is the critical driving factor behind the variation in emission line EW.

Combining the insights that (i) stronger metal emission line EW likely traces younger stellar populations, and (ii) greater metal emission line EW is associated with weaker feedback signatures (as demonstrated in Section~\ref{subsection:lya_vs_em}) leads to a picture in which feedback strengthens within the first $ \sim \SI{10}{Myr} $ of an episode of star formation as feedback from core-collapse SNe increases.  We discuss this prospect further in Section \ref{section:discussion}.

\subsection{Expanding Shell Models}
\label{section:zelda}

\subsubsection{Fitting and Quality Control}
\label{subsection:shell_fitting}

\begin{deluxetable}{ccc}
	\tablecaption{Summary of all parameters fitted by \texttt{zELDA}, including (i) parameter name, (ii) scale (log or linear), and (iii) uniform prior range.  }
	\tablehead{
		\colhead{Parameter} & \colhead{Scale} & \colhead{Prior range}
	}	 
	\startdata
	$ z_{\text{sys}} $  & linear & \emph{variable} \\
	$ V_{\text{exp}} $  & log    & $ [0.0, 3.0] $ \\
	$ N_{\text{H}} $    & log    & $ [17.0, 21.5] $ \\
	$ \tau_{\text{a}} $ & log    & $ [-4.0, 0.0] $ \\
	$ W_{\text{int}} $   & linear & $ [0.01, 6.0] $ \\
	$ EW_{\text{int}} $  & log    & $ [-1.0, 3.0] $ \\
	\enddata
\end{deluxetable}
\label{table:lyamcmcpriors}

As we showed in Section \ref{subsection:lya_vs_vout}, the relationship between metal absorption line profiles and emergent \lya profiles agrees qualitatively with expectations of expanding shell RT models.  This naturally raises the question of whether such models can directly be used to infer outflow properties from \lya profiles alone, without the need for detections of faint metal emission and absorption lines.  While such a prospect would greatly facilitate the study of these objects, as we will show, we find that simple expanding shell RT models are only partially successful in their ability to reproduce the relationships between \lya and outflow properties explored in Section~\ref{subsection:lya_vs_vout}.

\begin{figure*}[htbp]
	\centering
	\begin{tikzpicture}
		\node[anchor=south west] (imageA) at (0,0) {\includegraphics[width=0.31\linewidth]{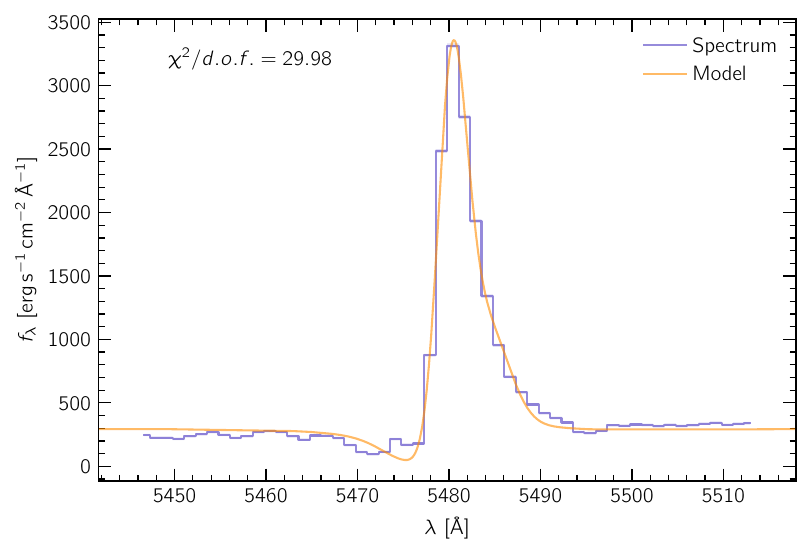}};
		\node[anchor=south west, xshift=25pt, yshift=-35pt] at (imageA.north west) {\small\textbf{(a)}};
	\end{tikzpicture}%
	\hfill
	\begin{tikzpicture}
		\node[anchor=south west] (imageB) at (0,0) {\includegraphics[width=0.31\linewidth]{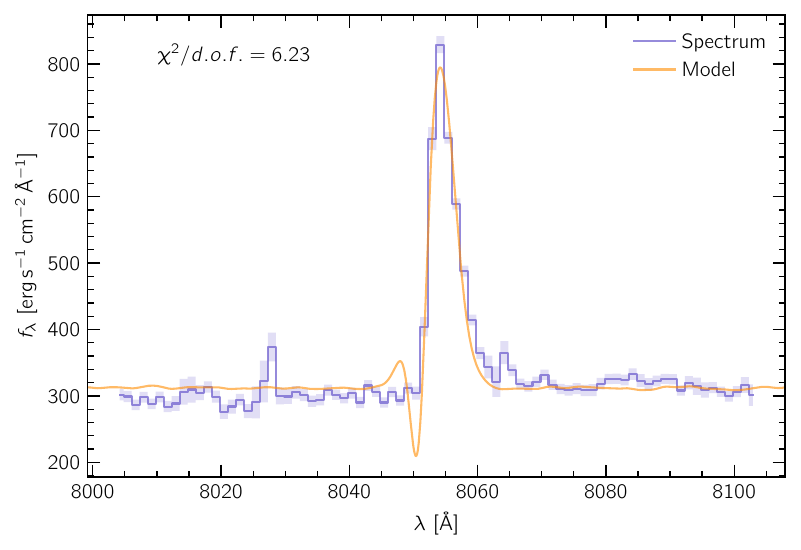}};
		\node[anchor=south west, xshift=25pt, yshift=-35pt] at (imageB.north west) {\small\textbf{(b)}};
	\end{tikzpicture}%
	\hfill
	\begin{tikzpicture}
		\node[anchor=south west] (imageC) at (0,0) {\includegraphics[width=0.31\linewidth]{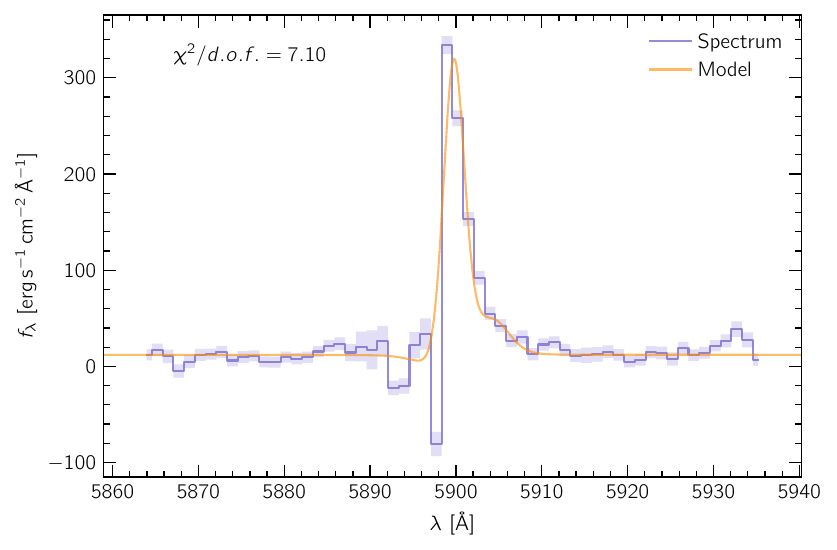}};
		\node[anchor=south west, xshift=25pt, yshift=-35pt] at (imageC.north west) {\small\textbf{(c)}};
	\end{tikzpicture}%
	\caption{Examples of \lya profiles with poorly-fitting expanding shell models. \textbf{(a)} a case in which the fit is qualitatively reasonable, but extreme SNR renders even small discrepancies highly statistically significant.  \textbf{(b)} a case in which the model is qualitatively incorrect: in this case including a spurious blueshifted peak as well as an absorption trough. \textbf{(c)} a case affected by residual sky lines (note the extreme negative flux density in the channel just short of the \lya line).}
	\label{fig:zeldafits}
\end{figure*}

To assess whether we can reliably infer outflow properties directly from the \lya profiles, we used the RT modeling code \texttt{zELDA} \citep{gl22, gl19}.  The latter provides routines to fit expanding shell RT models to \lya profiles and surrounding continuum via MCMC.  The grid of expanding shell models used by \texttt{zELDA} (described fully in \citealt{orsi12}) assumes a simple geometry consisting of a thin expanding shell (non-zero thickness, but thin relative to the distance from the central source) consisting of a homogeneous medium (i.e. uniform $ n_H $ everywhere).  Henceforth, we refer to this geometry as a homogeneous thin shell (HTS).  The incident \lya from the central source (i.e. prior to resonance scattering) is assumed to have a Gaussian profile.  Thus the model allows us to fit for: (i) three outflow properties, including expansion velocity $ V_{\text{exp}} $, \ion{H}{1} column density $ N_{\text{H}} $, and dust optical depth $ \tau_{\text{a}} $; (ii) the intrinsic \lya line profile in terms of FWHM $ W_{\text{int}} $, equivalent width $ EW_{\text{int}} $, and systemic redshift $ z_{\text{sys}} $.  We fitted only those \lya lines with sufficient SNR to yield reasonably well-constrained parameters, setting a minimum integrated SNR of $ 10 $ (giving us $ 374 $ spectra overall).  For each \lya profile, we input to \texttt{zELDA} a sub-spectrum of width $ \SI{15}{\text{\AA}} $ (in rest frame) centered on the \lya peak, adopting priors as shown in Table \ref{table:lyamcmcpriors}.  We take the median of the resulting posterior distributions to be the ``best-fit", and use the $ 16^{\text{th}} $ and $ 84^{\text{th}} $ percentiles to compute approximate $ 1\sigma $ uncertainties.

We find reasonable agreement between the best-fit models produced by \texttt{zELDA} and the observed spectra, with a median reduced-$ \chi^2 $ of $ 2.60 $ and $ 202/374 $ cases with reduced-$ \chi^2 < 3 $.  We break the 172 poor fits with reduced-$ \chi^2 > 3 $ down into three loosely-defined categories:

\begin{enumerate}
	\item Cases where the \lya line is qualitatively well-reproduced, but due to extreme SNR the deviation between the model and spectrum is highly significant ($ \sim 15 $ cases\footnote{The number of sources in each category is approximate due to the inherent subjectivity of our classification scheme}).  An example is shown in Figure \ref{fig:zeldafits}(a).
	\item Cases where the model fails to qualitatively reproduce the spectrum -- this happens either when there is a blue \lya peak which the model does not reproduce, when the model includes a blue peak not seen in the actual spectrum, and/or when the model includes a blueshifted \emph{absorption} component not seen in the spectrum ($ \sim 15 $ cases).  An example is shown in Figure \ref{fig:zeldafits}(b).
	\item Cases affected by systematics -- significant fluctuations in the spectrum beyond the spectral uncertainties, likely due to sky lines or residual contamination from nearby cluster members ($ \sim 140 $ cases).  An example is shown in Figure \ref{fig:zeldafits}(c).
\end{enumerate}

Cases in category 1 demonstrate that, when SNR is sufficiently high, deviations between HTS models and real \lya profiles can become apparent, as may be expected for any idealized model.  It is worth noting here that there are also 3 cases with SNR $ > 100 $ among the good fits with reduced-$ \chi^2 < 3 $, showing that some \lya profiles are reproduced by HTS models with remarkably high precision.  Cases in category 2 are also unusual in that they tend to have high \lya SNR suggesting that they are extreme instances of category 1, such that the discrepancies between model and data become appreciable even upon visual inspection.  Categories 1 and 2 demonstrate that real \lya profiles sometimes exhibit shapes that HTS models are unable to fully reproduce, though we note that such cases are relatively rare (just $ \sim 30 $ out of a sample of $ 374 $).  On the other hand, cases in category 3 mostly appear to reproduce the \lya profiles well based on visual inspection, with most of the systematics causing the high reduced-$\chi^2$ affecting the surrounding baseline rather than the \lya profile itself.  Nevertheless, out of an abundance of caution, we exclude sources from category 3 along with categories 1 and 2 from further analysis, leaving us with $ 202 $ fitted \lya profiles for which reduced-$ \chi^2 < 3 $.

\subsubsection{HTS Models vs Outflow Properties}
\label{subsection:hts_vs_outflows}

\begin{figure*}[ht!]
	\centering
	\includegraphics[width=\textwidth]{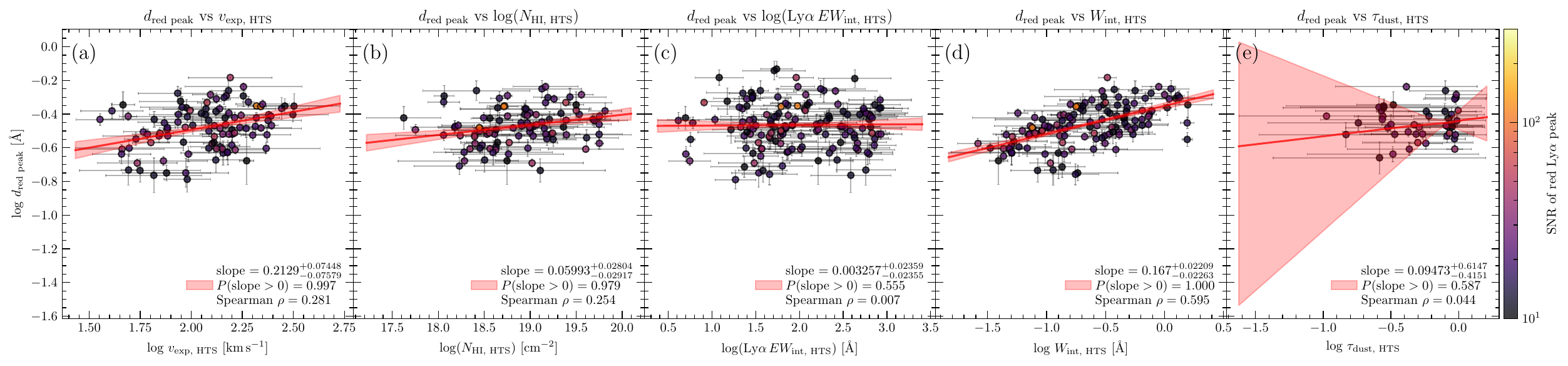}
	\caption{\lya dispersion plotted against fitted expanding shell model parameters from \texttt{zELDA}: (a) expansion velocity $ V_{\text{exp}} $, (b) neutral column density $ \log N_{\text{H}} $, (c) intrinsic \lya equivalent width $ EW_{\text{int}} $, (d) intrinsic \lya FWHM $ W_{\text{int}} $, and (e) dust optical depth $ \log \tau_{\text{a}} $. Fitted linear regression models produced by \texttt{LinMix} are shown in red.}
	\label{fig:zelda_vs_dispr}
\end{figure*}

We now evaluate the success of the fitted HTS models in capturing outflow properties. As shown in Section~\ref{subsection:lya_vs_vout}, we find \lya red peak dispersion $ d_{\text{red peak}} $ shows a correlation with outflow velocity measured via the centroids of blueshifted interstellar absorption lines. To see whether our HTS models capture this relationship, we compared the fitted expansion velocities $ V_{\mathrm{exp}} $ with measured \lya dispersion and fitted regression models using \texttt{LinMix} as shown in Figure~\ref{fig:zelda_vs_dispr}(a).  We also show how other fitted HTS parameters respond to $ d_{\text{red peak}} $ in panels (b)--(e). We now address each parameter in turn.

$ V_{\text{exp}} $ (panel (a)): the model expansion velocity correlates positively with $ d_{\text{red peak}} $ (Spearman $ \rho = 0.28 $), indicating that HTS models correctly identify broader \lya profiles as arising from faster expanding shells.  This qualitative agreement with the positive correlation between $ V_{\mathrm{out}} $ and $ d_{\text{red peak}} $ found in Section~\ref{subsection:lya_vs_vout} is encouraging in that it supports the use of $ d_{\text{red peak}} $ as a proxy for outflow velocity.  On the other hand, the scatter in the relationship is much larger than is suggested by the comparison of $ d_{\text{red peak}} $ and $ V_{\mathrm{out}}$ in Figure~\ref{fig:vout_dispr_corr}(a).  Furthermore, when we directly compared fitted values of $V_{\mathrm{exp}}$ with measured outflow velocities $V_{\mathrm{out}}$ we found no evidence of correlation between the two -- though given the small number of points (seven) we cannot rule out a correlation either.

\begin{figure}[ht!]
	\centering
	\includegraphics[width=0.45\textwidth]{"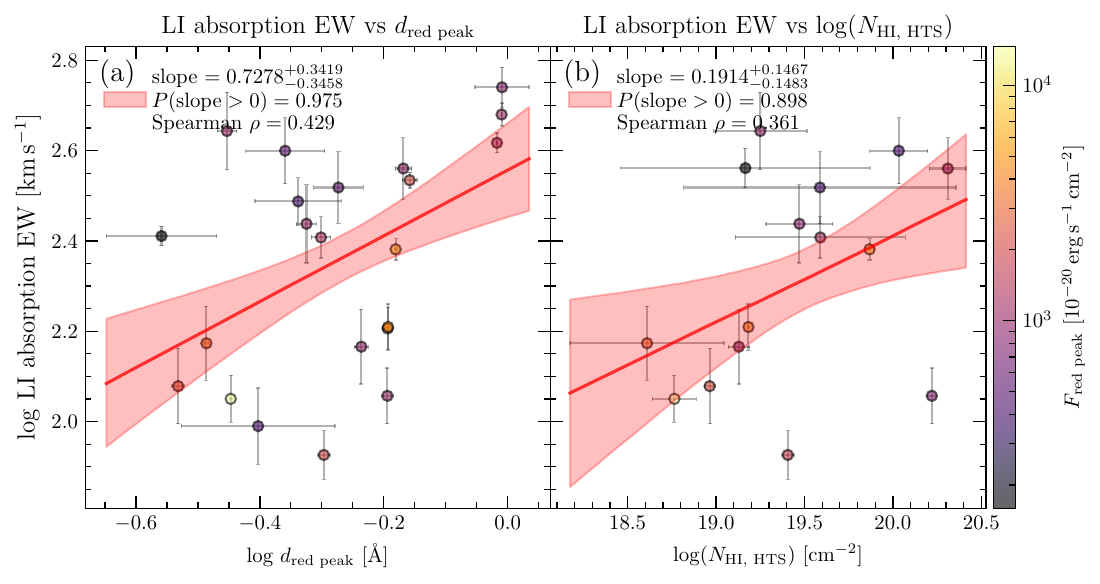"}
	\caption{Low-ionization interstellar absorption EW versus (a) \lya red peak dispersion and (b) neutral column density inferred using HTS models.}
	\label{fig:abs_ew_scatters}
\end{figure}

$ N_{\text{H}} $ (panel (b)): similarly to $ V_{\mathrm{exp}}$, the model neutral column density shows a tendency to increase with increasing $ d_{\text{red peak}} $ (Spearman $ \rho = 0.25 $).  This trend is unsurprising given that higher $ N_{\text{H}} $ increases resonant scattering through the expanding shell, broadening the emergent \lya profile.  As an empirical check on these trends, we plotted measured EW of the low-ionization absorption lines in our sample against both $ d_{\text{red peak}} $ and $ N_{\text{H}} $ and fitted regression models with \texttt{LinMix}.  The results, displayed in Figure~\ref{fig:abs_ew_scatters} show a tentative correlation between low-ionization absorption EW and $ d_{\text{red peak}} $ (panel (a)), indicating that the latter is indeed affected by neutral column density, in agreement with the HTS models.  However, the scatter is considerably greater than in Figure~\ref{fig:vout_dispr_corr}(a), suggesting a that neutral column density is a less powerful determinant of \lya dispersion than outflow velocity.  Comparing low-ionization absorption EW directly with fitted $ N_{\text{H}} $ in Figure~\ref{fig:abs_ew_scatters}, we find a plausible correlation, though the posterior probability of positive slope $ P(\mathrm{slope} > 0) $ is just $ \sim 0.9 $.  This result implies that HTS models are at least partially successful in estimating the relative neutral column density of the outflows in our sample.

$ EW_{\text{int}} $ (panel (c)): the intrinsic \lya EW fitted by \texttt{zELDA} shows no clear relation with $ d_{\text{red peak}} $, except for a very weak possible negative trend.

\begin{figure*}[ht!]
	\centering
	\includegraphics[width=0.75\textwidth]{"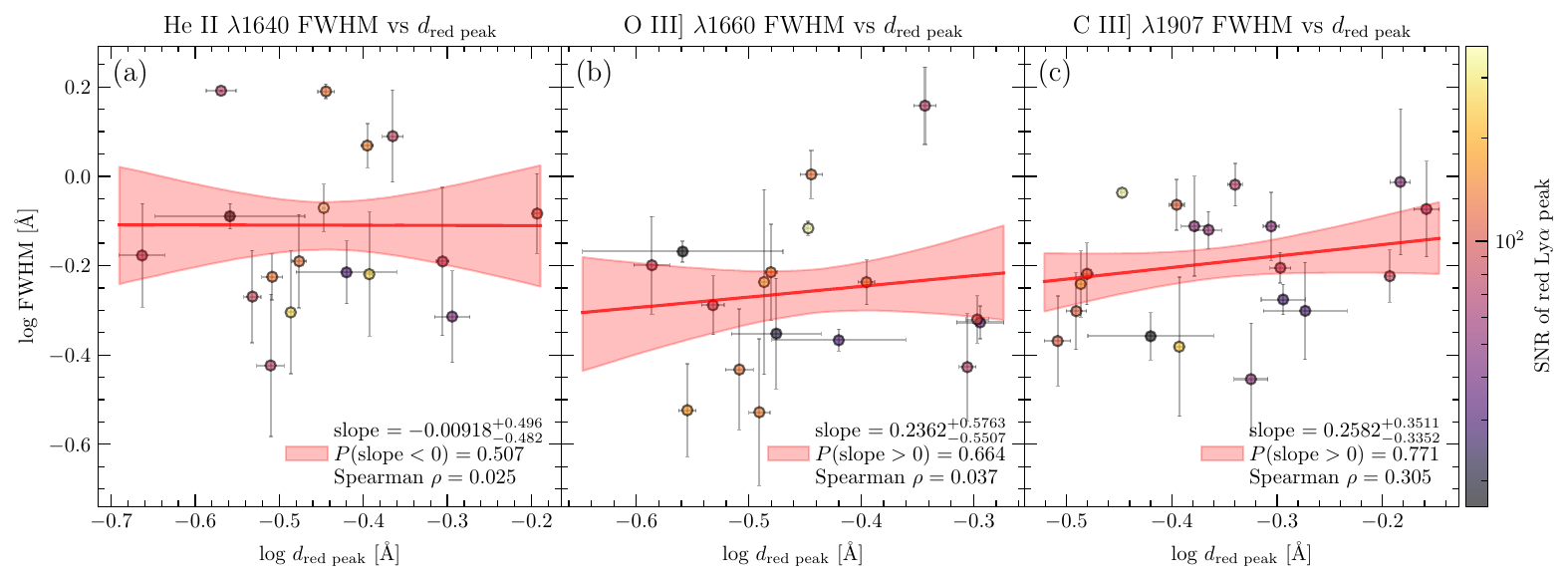"}
	\caption{FWHM of optically thin emission lines versus \lya dispersion.}
	\label{fig:em_fwhm_vs_dispr}
\end{figure*}

$ W_{\mathrm{int}} $ (panel (d)): the inferred intrinsic \lya line width based on HTS models shows a very strong correlation with measured \lya red peak dispersion (Spearman $\rho = 0.60$).  This is on the one hand unsurprising -- wider intrinsic lines should indeed be expected to produce wider emergent \lya profiles after radiative transfer -- however, it also raises the prospect that the the fitting over-relies on this parameter to explain variation in $ d_{\text{red peak}} $, at the expense of actual outflow properties.  To check whether this is the case, we plotted $ d_{\text{red peak}} $ against the width of the optically thin metal emission lines, as shown in Figure~\ref{fig:em_fwhm_vs_dispr}.  As these emission lines are not affected by scattering though the ISM/CGM, we can use them as a proxy for the intrinsic width of the \lya line.  We find no significant correlation between any of our optically thin emission lines and $ d_{\text{red peak}} $, consistent with the idea that the kinematics of outflowing gas -- rather than intrinsic \lya properties -- are the principal driver of the shape of the \lya line.

$ \tau_{\text{a}} $: we find no correlation between dust optical depth inferred from HTS models and \lya dispersion. We do, however, find that there is an unexpectedly large number of sources with $ \tau_{\text{a}} \sim 1 $, the upper bound of the range covered by the model grid; in most such cases $ \tau_{\text{a}} $ is poorly constrained.  However, in $ 5 $ cases $ \tau_{\text{a}} $ exceeds $ 0.1 $ by more than $ 2\sigma $, and manual inspection shows that all of these allegedly dusty sources have blueshifted \lya peaks.  This provides a crucial clue as to the origin of these unexpectedly high fitted $ \tau_{\text{a}} $ values, as we discuss later in this section.

\begin{figure}[ht!]
	\centering
	\includegraphics[width=0.45\textwidth]{"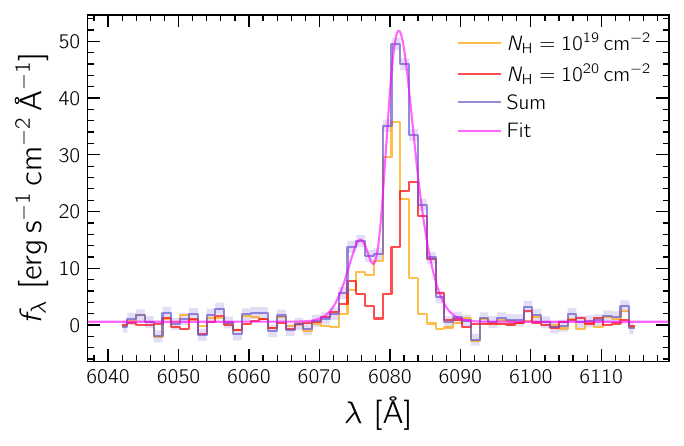"}
	\includegraphics[width=0.45\textwidth]{"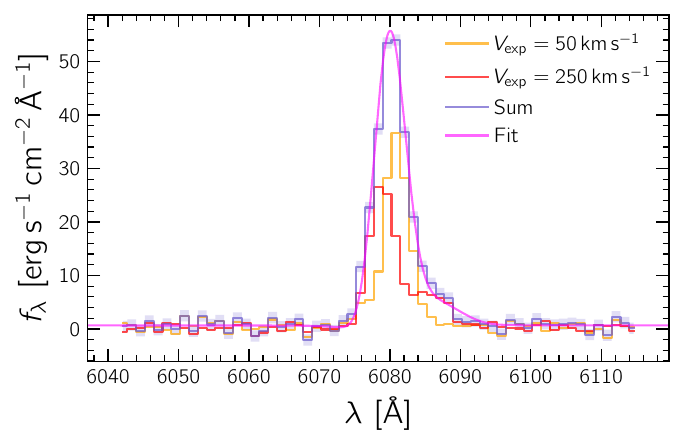"}
	\caption{\textit{Upper:} Simulated \lya profile for a clumpy expanding shell (shown in blue), approximated by summing together two HTS model profiles with different neutral column densities (shown in red and orange).  When fitted with a single HTS model, the dust optical depth $ \tau_{\text{a}} $ approaches $ 1 $, suggesting that a clumpy medium may cause homogeneous shell models to fit spurious high dust optical depth to real \lya profiles. \textit{Lower:} Simulated \lya profile for an outflow with more complex velocity structure, simulated by summing together two HTS models with different expansion velocities.  The fitted single HTS model here yields an under-estimated value of $ N_{\text{H}} $.}
	\label{fig:twocomp}
\end{figure}

\begin{deluxetable*}{ccccccccccc}
	\tablecaption{Input and output parameters used to approximate the effect of patchy outflows (rows 1 and 2) and complex velocity structure (rows 3 and 4) when fitting homogeneous expanding shell models.  Curly braces indicate where two HES \lya profiles of differing input values have been stacked together.}
	\tablehead{\colhead{$ V_{\text{exp,in}} $} & \colhead{$ \log N_{\text{H,in}} $} & \colhead{$ W_{\text{in,in}} $} & \colhead{$ \log EW_{\text{int}} $} & \colhead{$ \tau_{\text{a,in}} $} & \colhead{$ V_{\text{exp,out}} $} & \colhead{$ W_{\text{in,out}} $} & \colhead{$ \log EW_{\text{out}} $} & \colhead{$ \log N_{\text{H,out}} $} & \colhead{$ \tau_{\text{a,out}} $} & \colhead{Peaks}}
	\startdata
	20 & \{$ 19 $, $ 20 $\} & 0.1 & 2.0 & 0.01 & 89.8 & 0.39 & 2.11 & 18.05 & 0.9629 & DP \\
	150 & \{$ 19 $, $ 20 $\} & 0.1 & 2.0 & 0.01 & 249.9 & 0.43 & 1.93 & 19.43 & 0.0042 & SP \\
	\{$ 50 $, $ 100 $\} & $ 19 $ & 0.1 & 2.0 & 0.01 & 69.4 & 0.12 & 2.08 & 19.07 & 0.0291 & DP \\
	\{$ 50 $, $ 250 $\} & $ 19 $ & 0.1 & 2.0 & 0.01 & 281.4 & 0.16 & 2.18 & 18.55 & 0.0037 & SP \\
	\enddata
\end{deluxetable*}
\label{table:simfits}

As sources with $ z > 4 $ may have their \lya profiles significantly affected by IGM absorption shortwards of the \lya line (and this effect is not taken into account in the models used by \texttt{zELDA}), we re-fitted the above plots, this time using only sources with $ z < 4 $.  We found no qualitative differences from the trends shown in Figure \ref{fig:zelda_vs_dispr}, though posterior probabilities fell slightly due to the smaller sample sizes.  This likely reflects that the majority of \lya profiles are dominated by the redshifted \lya peak which is not subject to IGM attenuation.

Taken together, these results suggest that, in many cases, our HTS models yield parameters that are unphysical.  To understand why this may be the case, we considered two of the most obvious limitations of HTS models: (i) their assumed homogeneity in $ N_{\text{H}} $, and (ii) the use of a single expansion velocity rather than a range as in, e.g., \cite{verhamme06}.  Our analysis of metal absorption features casts doubt on each of these assumptions.  Firstly, as demonstrated in Section \ref{section:abslines}, metal absorption lines show strong evidence for optically-thick clumps with limited covering fraction rather than a homogeneous medium.  Secondly, profiles of some of the absorption lines shown in Figure~\ref{fig:absorbers0} exhibit asymmetric shapes suggesting a complex velocity structure (see the panels for MACS0940 SP317 and SMACS2013 SP74).  To test whether these factors can lead to inaccurate fitted parameters, we constructed toy models consisting of \lya profiles generated by summing together simulated HTS profiles with (i) two different values of $ N_{\text{H}} $ to approximate patchy/clumpy outflows and (ii) two different values of $ V_{\text{exp}} $ to approximate more complex velocity structure.  For the mixed-$ N_{\text{H}} $ models we tested two values of $ V_{\text{exp}} $: a slow outflow speed of $\SI{20}{km\,s^{-1}}$, and a faster speed of $ \SI{150}{km\,s^{-1}} $. For the mixed-$ V_{\text{exp}} $ models we kept $ N_{\text{H}} $ the same and tested two different mixtures of $ V_{\text{exp}} $ -- one with high and one with low contrast -- to see whether the effects on the fitted parameters scale with the velocity range spanned by the outflows.  In all cases, we adopted fiducial values of $ \tau_{\text{a}} $, $ EW_{\text{int}} $, and $ W_{\text{int}} $ based on typical fitted model values; these and all other input values are given in Table \ref{table:simfits}.  Our simulated profiles were constructed to match the MUSE observations in terms of spectral resolution and wavelength binning, and we added Gaussian perturbations to each channel to simulate noise (scaled so as to give us a peak SNR of $ 30 $ in each \lya profile).  We then fitted each of these simulated profiles using single-component HTS models in exactly the same way as for the real \lya profiles.  As shown in the example in Figure~\ref{fig:twocomp}, the toy \lya profiles can be reproduced convincingly by just a single HTS model.  The output parameters, shown in Table \ref{table:simfits}, demonstrate that

\begin{itemize}
	\item[(i)] Fitting HTS models to patchy/clumpy outflows (i.e. the mixed-$ N_{\text{H}} $ profiles in rows 1 and 2 of Table~\ref{table:simfits}) may lead to systematic overestimation of intrinsic \lya width $ W_{\text{int}} $, with output values being $\sim 4$ times greater than the input values.  This agrees with the qualitative picture presented in Figure~\ref{fig:twocomp}, which shows that the combined \lya profile is significantly wider than either of the individual profiles.  This effect also demonstrates that HTS models can use unphysical values of $ W_{\text{int}} $ as a way to explain \lya line width, helping to explain the strong dependence we find between $ W_{\text{int}} $ and $ d_{\text{red peak}} $ in Figure~\ref{fig:zelda_vs_dispr}.
	\item[(ii)] Fitting HTS models to patchy/clumpy outflows may lead to systematic overestimation of expansion velocity, while simultaneously underestimating the relative difference between slow and fast outflows -- $ V_{\text{exp}} $ increases by a factor of 4.5 for the slow outflow model, but by a factor of just 1.7 for the fast outflow model, thus reducing the contrast between the two.  This may help to explain why our fitted HTS models show only a weak dependence of $ V_{\text{exp}} $ on $d_{\text{red peak}}$.
	\item[(iii)] At the same time, fitting HTS models to patchy/clumpy outflows tends to exaggerate the role of $ N_{\text{H}} $ in determining the shape of the \lya profile, with slower and faster outflows being fitted with lower and higher values of $ N_{\text{H}} $, respectively.
	\item[(iv)] \lya profiles arising from clumpy outflows with low $ V_{\text{exp}} $ may indeed prefer extreme values of $ \tau_{\text{a}} $ when fitted with HTS models, but the same is not generally true for high-$ V_{\text{exp}} $ sources.
	\item[(v)] Fitting HTS models to profiles arising from more complex velocity structures (i.e. the mixed-$ V_{\mathrm{exp}} $ models in rows 3 and 4 of Table~\ref{table:simfits}) produces surprisingly accurate results, so long as the range of velocities spanned by the outflow is moderate.  On the other hand, when the profile is composed of multiple components with high velocity contrast, $ V_{\mathrm{exp}} $ becomes overestimated, while $ N_{\mathrm{H}} $ becomes underestimated.
\end{itemize} 

While our toy models go some way towards understanding what factors may feasibly underpin the failure of HTS models to accurately reproduce outflow properties of real \lya-emitting galaxies, some questions still remain.  Firstly, the mixed-$ N_{\text{H}} $ model leads to a dramatically \emph{underestimated} value of $ N_{\text{H}} $ in the low-$ V_{\text{exp}} $ case but not in the high-$V_{\mathrm{exp}}$ case -- an effect that we found to be persistent across a wide range of input parameters.  Secondly, faster outflow models (rows 2 and 4 in Table~\ref{table:simfits}) are fitted with underestimated dust optical depth.  That discrepancies remain is unsurprising, as our simple models neglect any cross-talk between different components, which are likely to have a significant impact on the emergent \lya profile (e.g. if outflow velocity depends on radius, then \lya photons emerging from slower inner regions must undergo radiative transfer through faster outer regions).  Furthermore, real outflows are likely to be better represented by a continuous range in both $ N_{\text{H}} $ and $ V_{\text{exp}} $ rather than two distinct values each.  A systematic investigation of \lya RT is beyond our current scope; our toy models merely serve to show that deviations from simple HTS geometry can lead to unphysical fitted model parameters in line with what we find when we fit HTS models to real \lya profiles.

\begin{figure}[ht!]
	\includegraphics[width=0.47\textwidth]{"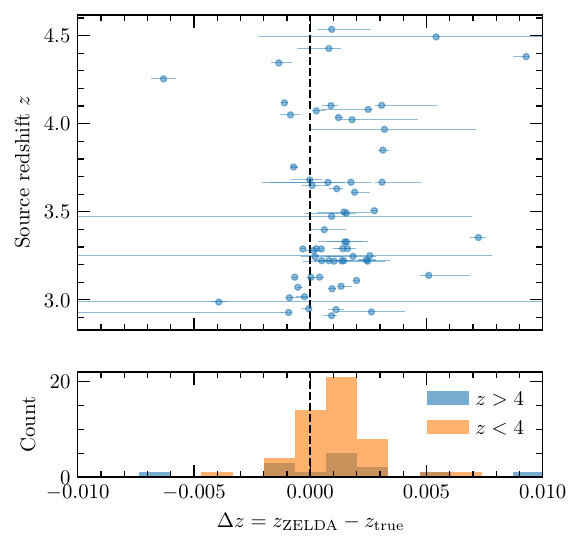"}
	\caption{Comparison of estimated redshift from expanding shell models with that measured using optically thin emission lines.  Uncertainties are shown at the $ 1\sigma $ level. In the histogram (lower panel) the sources are divided between those with $z > 4$ and $z < 4$. The negative skew of the distribution in $ z_{\text{zELDA}} - z_{\text{true}} $ indicates that HTS models tend to overestimate the systemic redshift.}
	\label{fig:zeldazcheck}
\end{figure}

We also checked the accuracy of the \texttt{zELDA} estimates of $ z_{\text{sys}} $ by comparing them against the systemic redshift as measured using optically-thin emission lines.  Figure \ref{fig:zeldazcheck} compares the systemic redshift measured from these emission lines with that from \texttt{zELDA} in velocity space (relative to the red \lya peak).  In general, the differences between the two values are within $ 3\sigma $ of zero; however, the distribution of redshift discrepancies, shown in the histogram in the lower panel, is centered above zero.  This means that expanding shell models tend to underestimate the velocity offset of the red \lya peak relative to the systemic (i.e. placing the systemic too close to the red \lya peak in velocity space).  This raises three major potential problems:

\begin{itemize}
	\item[(i)] The inferred outflow kinematics may be inaccurate, since the proximity of the redshifted peak relative to the systemic scales with the expansion velocity (e.g. \citealt{verhamme06}).
	\item[(ii)] If there are also metal absorption lines present, their blueshift relative to the systemic velocity will be exaggerated, leading to overestimation of outflow velocities.
	\item[(iii)] The amount of \lya emission at the systemic velocity may be overestimated.  This has important consequences, as it has been found that -- at low redshift -- systemic \lya emission can be used as a proxy for LyC escape \citep{verhamme15}.  Any future studies employing this relationship and wishing to use expanding shell models to estimate systemic redshifts may consequently overestimate LyC leakage.
\end{itemize}

The same over-estimation of systemic redshift has also been reported by \cite{orlitova18}, who fit similar expanding shell models to low-redshift Green Pea galaxies.

\section{Discussion}
\label{section:discussion}

\subsection{Evidence of Delayed SNe Feedback}
\label{subsection:snefeedback}

The analysis in Sections~\ref{subsection:lya_vs_vout}, \ref{subsection:lya_vs_em}, and \ref{subsection:ew_factors} points to an intimate connection between stellar evolution and feedback in the first $\sim \SI{10}{Myr}$ following the onset of star formation, with the very youngest sources showing slower outflow speeds.  We interpret this as evidence of strengthening feedback over the first few Myr of star formation, the timescale aligning well with expectations of the onset of core-collapse SNe feedback from the most massive stars after $ \sim \SI{4}{Myr} $ (see \citealt{woosley02,krumholz14} for details of the relevant stellar evolutionary timescales, \citealt{agertz13, hopkins23} for implementation into galaxy simulations). Recent studies of young star clusters in nearby galaxies with JWST NIRCam and MIRI find a comparable time scale of $ \sim \SI{3}{Myr} $ for the disruption of photo-dissociation regions by stellar feedback \citep{pedrini24}.  Of course, such young clusters are far less massive than the galaxies in our sample ($ 10^3 $ -- $ \SI{10^4}{M_\odot} $ vs $ \SI{10^6}{M_\odot} $ or more).  However, the contrast in physical size is much less -- the nearby young star clusters can be a few tens of pc in size, similar to some of the smaller objects in our sample (see \citealt{claeyssens22}) -- suggesting that feedback timescales may also be similar.

\begin{figure}[ht!]
	\includegraphics[width=0.47\textwidth]{"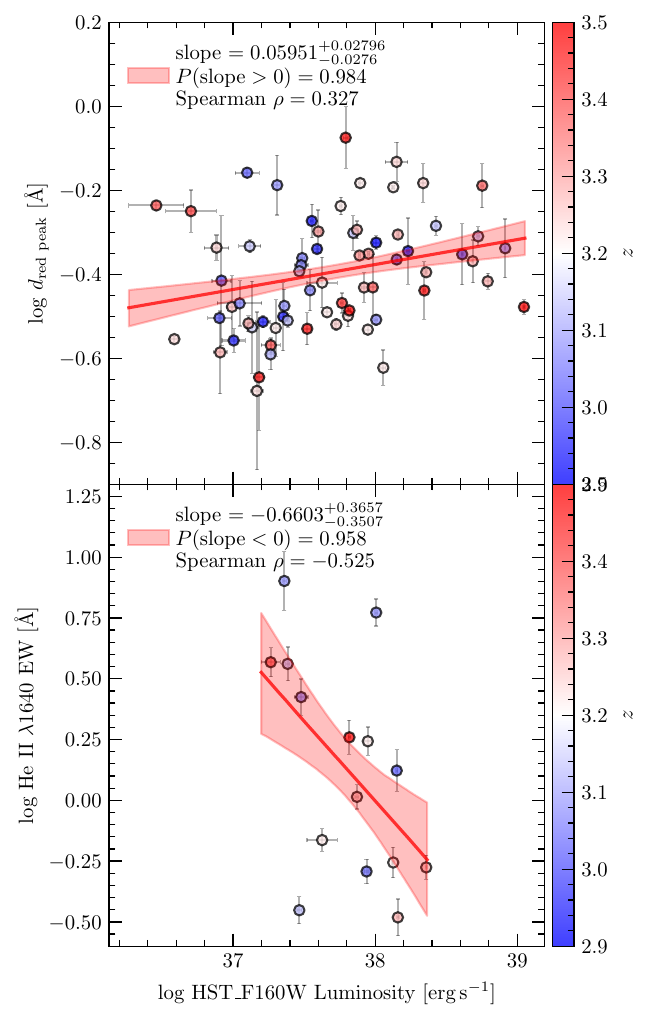"}
	\caption{Upper: \lya dispersion plotted against HST F160W luminosity (taken to be a rough proxy for stellar mass), showing a tentative increasing trend. Lower: \heii EW plotted against HST F160W luminosity, showing a tentative decreasing trend.}
	\label{fig:ew_mass_corr}
\end{figure}

Another factor that may influence the relationship between age and outflow rate is galaxy mass. Lower-mass SFGs will tend to drive slower outflows due to lower escape velocities \citep{thompson24}, and yet may also undergo more bursty star formation \citep{benavides25} and hence have younger average ages when selected by \lya emission or UV luminosity.  Of course, this effect is also deeply connected to feedback timescales, with the shortest feasible burst being set by the age at which SNe feedback kicks in.  In the extreme scenario in which SNe feedback halts star formation completely, star formation episodes in low-mass sources may simply not survive past $ \sim \SI{4}{Myr} $.  This would mean that sources with lower mass suffer sharp declines in their emission line EWs, while more massive sustained starbursts would spend more time in ``low-emission-EW mode", potentially explaining the relation between outflow velocity and emission line EW without needing to invoke a scenario where feedback strengthens as the stellar population ages.  To check whether this may be the case, we plotted \lya dispersion $ d_{\text{red peak}} $ and \heii EW against luminosity in the HST F160W filter provided in the \rto catalogs, using the latter as a very rough proxy for stellar mass.  The F160W filter covers rest-frame near-UV/optical ($ 3500 $ -- $ \SI{4000}{\AA} $) at redshifts between 3 and 3.5 and hence is less sensitive to the instantaneous SFR than the shorter-wavelength HST filters.  We restrict sources in the plot to this redshift range to minimize the impact of variability in the rest-frame wavelength range covered by the F160W filter.  The result, shown in Figure~\ref{fig:ew_mass_corr}, shows a tentative correlation between $ d_{\text{red peak}} $ and F160W luminosity, as well as a tentative anticorrelation between \heii EW and F160W luminosity.  Based on these results, we consider it plausible that stellar mass may help at least in part to drive the relationship between stellar population age and outflow velocity.  However,  we caution that (i) rest-frame near-UV/optical luminosity alone is limited as a gauge of overall stellar mass and (ii) older, more sustained starbursts may have more time to build up stellar mass, which may drive the correlations seen in Figure~\ref{fig:ew_mass_corr} even under a scenario where feedback strengthens with age. Robust mass estimates and star formation histories from joint HST and JWST photometry will likely shed more light on this issue, which we defer to a future paper.

We also considered the impact of line-of-sight effects on the emergent \lya profile.  Radiation-hydrodynamic simulation of a \lya emitting galaxy of $ M \sim \SI{10^9}{M_{\odot}} $ by \cite{blaizot23} show that the observed \lya profile can vary dramatically depending on viewing angle due to anisotropic neutral column density/ISM kinematics.  While viewing angle likely contributes considerable scatter to the relationship between age and interstellar absorption line profiles, its effect may be mitigated somewhat by the generally low masses of our target sources ($ \sim 10^6 - \SI{10^9}{M_{\odot}} $; see, e.g., \citealt{iani24}), which may give rise to less collimated, more isotropic outflows (e.g. \citealt{strickland09}).

Comparable trends to those we find here in low-$z$ systems have been reported by \cite{hayes23i} and \cite{hayes23ii}, who find that (i) outflows accelerate as the driving stellar population ages, and (ii) stacked spectra of \lya emitting galaxies with more intense ionization conditions (as measured by the O$_{32}$ ratio) show greater \lya line widths as well as more strongly blueshifted low-ionization absorption lines with greater EW. Our findings support these results, and extend these trends to the early universe.

\section{Summary and Conclusion}
\label{section:sum}

We analyzed the spectra of 573 gravitationally lensed \lya-emitting sources at $ 2.9 < z < 6.7 $, fitting analytic line profiles to \lya and rest-frame UV metal emission/absorption features.  We then used Bayesian linear regression (\texttt{LinMix}) and comparison with stellar population + photoionization model grids \citep{gutkin16} to connect \lya morphology, outflow kinematics, and stellar population age.  Our main conclusions are as follows:

\begin{itemize}
	\item[(i)] These sources show spectral signatures suggesting near ubiquitous outflows.  Single \lya peaks are redshifted relative to the systemic velocity as measured using metal emission lines, or, when there is also a blue peak, the two peaks straddle the systemic.  Metal absorption lines from both low- and high-ionization species are blueshifted (or consistent with blueshifted given uncertainties) relative to the systemic redshift in most cases where the latter can be measured (18/22).
	\item[(ii)] Both low- and high-ionization absorption lines exhibit line ratios indicative of saturation, suggesting that the outflows consist of optically thick clumps of neutral gas as well as warm ionized gas with limited covering fractions.
	\item[(iii)] Outflow velocity inferred from low-ionization absorption lines correlates positively with \lya intrinsic dispersion, establishing that \lya profile width traces genuine outflow kinematics.  This provides a much needed empirical basis for interpreting \lya profiles in terms of outflow properties.
	\item[(iv)] UV metal emission-line EWs, especially \heii and \civ, are anti-correlated with \lya width/offset diagnostics (intrinsic dispersion, FWHM, and for double-peaked sources the blue-red separation), and show tentative direct anti-correlation with absorption-line outflow velocity.  Because stronger metal emission traces younger, harder ionizing populations, these trends imply that the youngest starbursts have narrower \lya profiles and weaker/slower outflows.
	\item[(v)] Comparison with \citep{gutkin16} model grids indicates that sources with the strongest \heii emission are consistent with very young stellar populations, while sources with weak/undetected \heii are consistent with more evolved populations.  Taken together with points (iii)--(iv), this supports a delayed-feedback scenario in which outflows strengthen over the first $ \sim \SI{10}{Myr} $ after the onset of star formation, as feedback from supernovae increases.
	\item[(vi)] Homogeneous thin-shell RT fits (\texttt{zELDA}) often reproduce \lya line shapes but can return unphysical parameters and systematically overestimate systemic redshifts.  Consequently, such models should not be used uncritically to infer outflow properties without external constraints.
\end{itemize}

Our most salient result -- that stellar feedback appears to strengthen dramatically with stellar age within the first $ \sim \SI{10}{Myr} $ of an episode of star formation -- provides a critical empirical constraint on feedback timescales in young galaxies. This timescale, likely tied to the onset of core-collapse supernovae, has significant implications for models of early galaxy formation, where the efficiency and timing of feedback govern predictions for star formation, gas ejection, and chemical enrichment.  While our findings suggest a feedback-driven transition in outflow properties, two scenarios remain plausible: (i) the observed correlations reflect a temporal sequence, in which star-forming galaxies develop progressively stronger outflows over the first $ \sim \SI{10}{Myr} $ as SNe feedback develops, manifesting as progressively broader \lya profiles and larger absorption-line outflow velocities at fixed stellar mass; or (ii) the correlations primarily reflect a mass sequence, in which more massive galaxies sustain longer episodes of star formation, drive faster outflows, and display broader \lya profiles.  JWST imaging of these lensing clusters -- already available for several targets -- may soon resolve this degeneracy through robust stellar mass estimates.

\section{Acknowledgments}

	We thank Ishika Kaur and Wenjun Chen for their help in testing early versions of the spectral extraction and fitting routines used in this study, Johan Richard for his help accessing and understanding the data used in \rto, Masami Ouchi for his feedback on an early version of the manuscript, and the anonymous referee for greatly helping to improve the quality of the results and analysis herein.  J.N. and J.L. acknowledge support from the Research Grants Council of Hong Kong for conducting this work under the General Research Fund 1732122 and 173020323. J.N. acknowledges support from the Dissertation Year Fellowship issued by the University of Hong Kong.
	
	\software{Astropy \citep{astropy:2013, astropy:2018, astropy:2022}, Matplotlib \citep{mplref}, MPDAF \citep{mpdafref}, NumPy \citep{numpy}, Scikit-learn \citep{scikit-learn}, SciPy \citep{scipy}, zELDA \citep{gl22}}
	
\appendix

\section{\lya Profile Fitting Procedure}
\label{lya_fitting_appendix}

\subsection*{Emission model and initialisation}

Each \lya spectrum is fitted within a $\pm 50$\,\AA\ window centred on the red peak, with channels affected by sky-line residuals or bad pixels masked. A double-peaked fit is attempted first. To mitigate sensitivity to initial conditions, the blue-peak centroid is initialized at five offsets spanning $\pm 0.9(1+z)$\,\AA\ from its default starting position. For each initialization, a two-component asymmetric Gaussian is fitted using the Trust Region Reflective algorithm. A converged solution is accepted only if it simultaneously satisfies the following criteria: (i) both peaks are detected at S/N\,$>3$; (ii) neither S/N exceeds 5000 (flagging pathological solutions); (iii) the blue-peak amplitude does not exceed that of the red peak; (iv) the peaks are spectrally resolved; (v) the blue peak lies at shorter wavelength than the red peak; and (vi) their velocity separation is $<1000$\,km\,s$^{-1}$.

\begin{figure*}[ht!]
	\centering
	\includegraphics[width=0.95\textwidth]{"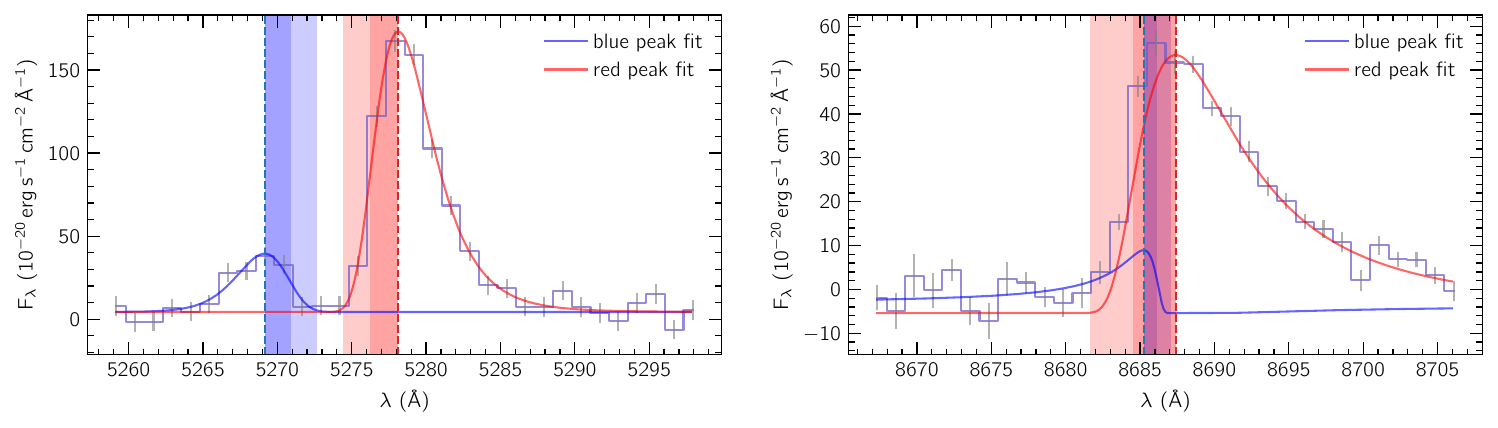"}
	\caption{\lya profiles demonstrating the criterion by which we consider the red and blue peaks to be resolved (left) or unresolved (right).  In both cases, the lines have been fitted with two asymmetric Gaussians.  The peak wavelengths are shown with dashed lines, while the dark shaded bands indicate the distance between the respective peaks and the points at which the profiles obtain half of their maximum values (the light shaded bands show double this distance).  For the two components to be considered resolved, neither of the shaded bands of one component may overlap with the peak wavelength of the other component.  In the resolved case, the blue band does not extend to longer wavelengths than the red dashed line, nor does the red band extend to shorter wavelengths than the blue dashed line.  In the unresolved case, the red band covers the blue dashed line.}
	\label{fig:resolved_lya}
\end{figure*}

Spectral resolution of two asymmetric Gaussians B and R is defined by the criterion
\begin{equation}
    \lambda_{\text{c,R}} - \lambda_{\text{c,B}} > \max(W_{\text{r,B}},\, W_{\text{l,R}})
\end{equation}
where $W_{\text{r,B}}$ ($W_{\text{l,R}}$) is twice the distance from the peak of B (R) to the wavelength at which it reaches half its maximum value on its red (blue) side; see Figure~\ref{fig:resolved_lya} for illustrative examples.

If no solution passes all criteria but a solution fails only the resolution criterion, a single further attempt is made in which the asymmetry parameter of the blue peak, $a_{\rm blue}$, is constrained to be negative (left-skewed, $a_{\rm blue} < -0.1$), using the partially-passing solution as the starting point. This is physically motivated: a resolved blue peak should be skewed towards the red peak. If this also fails, the spectrum is described by a single asymmetric Gaussian.

\subsection*{Baseline and peak-number selection}

All three baseline models (constant, linear, DLA) are evaluated for each source. The preferred combination of baseline and peak number (single or double) is determined by the Bayesian Information Criterion,
\begin{equation}
    \mathrm{BIC} = \chi^2 + k \ln n
\end{equation}
where $\chi^2$ is the total chi-squared of the fit, $k$ is the number of free parameters, and $n$ is the number of fitted spectral channels. The constant baseline has 5 (single peak) or 9 (double peak) free parameters; the linear baseline adds one additional parameter (slope); the DLA baseline adds three (optical depth $\tau$, FWHM, and line centre of the Voigt profile). A DLA solution is rejected if the fitted continuum level is non-positive. The model with the lowest BIC is selected, with a full bootstrap run (1000 iterations) performed only for the winning model to avoid unnecessary computation. During bootstrapping, spectral uncertainties blueward of the red peak are inflated by the median ratio of fit residuals to formal uncertainties measured in that region (using a cubic polynomial baseline with $3\sigma$ clipping to identify the continuum), helping to account for excess variance caused by Ly$\alpha$ forest absorption.

\section{Other Emission and Absorption Line Fitting}
\label{app:other_fitting}

For emission and absorption lines other than \lya, we fitted a conventional Gaussian profile together with a linear function for the local continuum within $\pm\SI{25}{\text{\AA}}$ of the reported observed wavelength in the \rto catalogs.  In the case of doublets such as \ciii, we fitted two Gaussian profiles simultaneously with the same central velocity and width, leaving the respective amplitudes as independent parameters.  In addition, we forced fits to a set of key diagnostic lines --- namely \civ, \siii, \oiii, \siiv, \ciii, and \heii --- whenever they fell within the spectral coverage of the extracted spectrum but were absent from the \rto catalog for that source, using the \lya redshift to construct initial guesses.  We constrained the central wavelengths of the Gaussians to be within $\SI{\pm1.5625}{\text{\AA}}$ of the expected rest-frame wavelength (equivalent to ${\approx}\SI{6.25}{\text{\AA}}$ in the observed frame at $z \approx 3$, or approximately five spectral channels of MUSE).  We also constrained the FWHM of the fitted Gaussians to be at least $\SI{2.0}{\text{\AA}}$ in wavelength space and at most $\SI{300}{km\,s^{-1}}$ in velocity space: the lower bound reflects the approximate spectral resolution of MUSE and helps to avoid fitting narrow features that are occasionally left after sky subtraction, while the upper bound prevents fitting gradual baseline fluctuations that can arise due to contamination by nearby cluster members.  Uncertainties in all fitted parameters were obtained directly from the covariance matrix produced by the fitting module.  This approach was preferred over bootstrap resampling because of its much lower computational cost (requiring only a single evaluation of the fitting routine per line) and the large number of emission/absorption lines in the sample ($\sim 20\,000$).  As covariance-matrix uncertainties tend to be underestimated, any putative $\geq 3\sigma$ detections are considered tentative only and followed up with more robust bootstrap fitting as needed.

\section{Quality-Control Procedures}
\label{app:qc}

\subsection*{\lya outlier removal}

We searched for extreme outliers among the \lya line profiles using the line asymmetry and flux uncertainty (of the more redshifted peak in cases where two peaks were fitted), along with the surrounding continuum level.  The asymmetry and flux uncertainty are useful in finding \lya lines strongly affected by systematics due to residual atmospheric features and/or contamination by emission/absorption from foreground galaxies, while the continuum level helps to identify spectra strongly contaminated by foreground continuum sources.  We assessed the severity of outliers by ranking them according to their Local Outlier Factor \citep[LOF; see][]{lof_paper} based on the 5 nearest neighboring points in parameter space, and manually selected a cutoff above which \lya profiles were judged to be dominated by systematics.  This removed 33 spectra, leaving 1052.

\subsection*{Foreground contamination flagging}

As the target sources are situated behind galaxy clusters, there is a danger that spectral features from foreground objects can become imprinted on the target spectra and mistaken for intrinsic features.  Although \rto perform local background subtraction on the source spectra, manual inspection reveals some cases where residual features from foreground galaxies such as the \ion{Ca}{1} H and K lines are visible in the target spectra.  To identify such cases, we first searched the \rto catalogs for foreground stars, cluster members, and/or intermediate redshift galaxies with redshifts below 2.9 (i.e.\ below all \lya emitters in our sample) whose centroids fall within $10.0\arcsec$ of each source.  We then identified which of these sources are likely to have the largest impact on the target spectrum by computing a contamination score for each source,
\begin{equation}\label{equation:distmod_app}
	S \equiv \frac{F_c}{R_{\text{ISO}}} e^{\left(- D / R_{\text{ISO}}\right)}
\end{equation}
where $F_c$ is the flux contained within an aperture of diameter 1 PSF FWHM at the source center, $R_{\text{ISO}}$ is the radius inferred from the isophotal area listed in the \rto catalog, and $D$ is the angular separation from the target in arcseconds.  We found this score to provide the best balance between distant but extended sources and close but compact sources in testing.  Having established the most likely contaminating object, we estimated the degree to which it actually contaminates the source.  To do this, we generated a 2D image of the object from the MUSE data cube by averaging together all wavelength channels within $[5000, 7000]$\,\AA, and fitted it with a S\'{e}rsic profile.  We then used the intensity of the S\'{e}rsic profile at the position of the target source to estimate how much flux is contributed by the potential contaminating object within the aperture used to extract the target spectrum.  We used this value to normalise the spectrum of the potential contaminating object and hence estimate the degree of spectral contamination at the target position.  In the final step, we determined whether the potential contaminating object could be responsible for spurious line detections in the target source.  For each fitted line with SNR $> 3$ in the target spectrum, we attempted to fit the same line to the normalised contaminant spectrum, holding the central wavelength and the sign of the flux (i.e.\ absorption or emission) fixed, but allowing the amplitude and width of the fitted Gaussians to vary.  We used the spectral uncertainties of the target spectrum rather than the normalised contaminant spectrum; thus, if the degree of contamination was small, the target spectrum uncertainties were large compared to any spectral features in the contaminant, and no statistically significant fit was found.  If statistically significant fits were found, we flagged the corresponding line in the target spectrum as likely contaminated and excluded it from further analysis.  In total, 59 putative detections of spectral lines across 30 spectra were flagged via this procedure.

\section{Autoregression Model for Bootstrap Error Estimation}
\label{app:autocorrelation}

When fitting Gaussian profiles to search for absorption lines, we found that a significant fraction of spectra exhibit correlated noise -- likely arising from imperfect sky subtraction -- which can produce spurious line detections. To account for this, we incorporated noise correlations into our bootstrap uncertainty estimation.

\subsection*{Estimating the correlation length}

For each spectrum, we first obtained a best-fit model using \texttt{scipy.curve\_fit} \citep{scipy} and computed the residuals. We then measured the autocorrelation function (ACF) of these residuals using an unbiased estimator,
\begin{equation}
    \mathrm{ACF}(k) = \frac{1}{n - k} \sum_{i=1}^{n-k} \epsilon_i \, \epsilon_{i+k}
\end{equation}
where $\epsilon_i$ is the residual in the $i^{\rm th}$ channel and $k$ is the lag in pixels, normalised so that $\mathrm{ACF}(0) = 1$. Significant correlation was declared at lag 1 if $\mathrm{ACF}(1) > 2/\sqrt{n}$, where $n$ is the number of spectral channels (a 2$\sigma$ criterion appropriate for short spectra of $30$--$100$ channels). If no significant correlation was detected, the noise was treated as uncorrelated.

Where significant correlation was detected, we estimated the e-folding correlation length $\tau$ by fitting an exponential decay,
\begin{equation}
    \mathrm{ACF}(k) = \exp\!\left(-\frac{k}{\tau}\right)
\end{equation}
to the ACF over the lags where it remained above a threshold of 0.05, using a weighted least-squares fit with zero intercept (enforcing $\mathrm{ACF}(0) = 1$). The fit was restricted to lags of $\lesssim 7$ pixels to avoid contamination by long-range baseline structure. In addition, we checked for systematic excess variance by comparing the median absolute residual to the formal spectral uncertainties; where this ratio exceeded unity, the uncertainties were inflated by that factor prior to noise generation.

\subsection*{Generating correlated bootstrap noise}

Bootstrap iterations were generated using an AR(1) (autoregression of order 1) process. Given correlation length $\tau$, the autoregressive coefficient is
\begin{equation}
    \phi = \exp\!\left(-\frac{1}{\tau}\right)
\end{equation}
and correlated noise realisations $\delta_i$ are drawn according to
\begin{equation}
    \delta_i = \phi \, \delta_{i-1} + \sqrt{1 - \phi^2} \, \xi_i
\end{equation}
where $\xi_i \sim \mathcal{N}(0, 1)$ are independent standard normal deviates. This construction ensures that the marginal variance of $\delta_i$ is unity and that the covariance structure is consistent with an exponential ACF with e-folding length $\tau$. The noise realisations are then scaled by the (possibly inflated) per-channel uncertainties $\sigma_i$ to produce the perturbation applied to the spectrum in each bootstrap iteration: $\Delta_i = \sigma_i \, \delta_i$. For spectra with no detected correlation ($\tau \leq 1$), standard independent Gaussian perturbations are used.

\section{Automated Line-Quality Flagging}
\label{app:flagging}

Following fitting, each detected line is assigned one or more single-character flags indicating potential quality issues.  The six flag types are described below.  Multiple flags may be combined for a single line.  As a special case, when both components of a doublet are independently detected at $|\mathrm{SNR}| \geq 3$ with the same flux sign and neither component carries a pre-existing contamination flag (\texttt{c}), all other flags are cleared: the simultaneous detection of both doublet lines at the correct wavelength ratio constitutes strong evidence for a genuine feature.

\textbf{Sky (\texttt{s}):} Sky-line contamination is assessed by checking whether the fitted peak wavelength falls within $\max(3\,\sigma_{\lambda_0}, 2.4\,\text{\AA})$ of a known MUSE sky line, where $\sigma_{\lambda_0}$ is the uncertainty on the fitted peak wavelength.

\textbf{Thin (\texttt{t}):} Lines whose FWHM (including its uncertainty) falls below the MUSE spectral resolution threshold of $2.05$\,\AA\ are flagged as unresolved, since such narrow features are more likely to arise from sky-subtraction residuals than from genuine emission or absorption.

\textbf{Negative (\texttt{n}):} Absorption lines for which more than half the spectral pixels within $\pm 2\,\mathrm{FWHM}$ of the peak are negative, or whose median flux in that window is negative, are flagged.  Such profiles generally arise due to residual sky lines left after sky subtraction.

\textbf{Peak-dominant (\texttt{p}):} A flag is raised when the single-pixel flux at the line centre (after continuum subtraction) accounts for more than 50\% of the integrated line flux.  This can indicate an unresolved noise spike rather than a spectrally extended emission or absorption feature.

\textbf{Multi-peak (\texttt{m}):} We check for unexplained features in the vicinity of the fitted lines by convolving the continuum-subtracted spectrum by a Gaussian with the same width as the fitted line, weighted by the inverse variance at each pixel. The resulting matched filter response array represents, at each wavelength position, the SNR of a Gaussian-shaped feature of the fitted width. The response at the fitted line centre is taken as the target score. A background distribution is then constructed from the response values at all other positions in the fitting window, excluding a guard region of $ \pm 1 \times $ FWHM around the target and any doublet partner or nearby unrelated spectral lines. We then fit a Gaussian to this background distribution via maximum likelihood, and the target score is evaluated against the resulting cumulative distribution function to yield a measure of its statistical significance.  To account for uncertainty in the fitted line's width, $ \sigma_{W} $, we repeat this procedure twice more, convolving with Gaussians of FWHM $ W_{\mathrm{fitted}} \pm \sigma_{W} $.  This approach accounts for the look-elsewhere effect: a line at formal $ 3 \sigma $ significance is only considered robust if its matched filter score is genuinely exceptional relative to the population of comparable features present across the same spectral window.

\textbf{Contaminated (\texttt{c}):} This flag is set by the foreground-contamination procedure described in Appendix~\ref{app:qc} and indicates that the line may be attributable to a foreground source rather than the target galaxy.  If this flag is already present when the automated flagging is run, it is preserved and the other flagging tests are skipped.

\section{Origin of Metal Emission Lines}
\label{app:emorigin}

\begin{figure}[ht!]
	\centering
	\includegraphics[width=0.47\textwidth]{"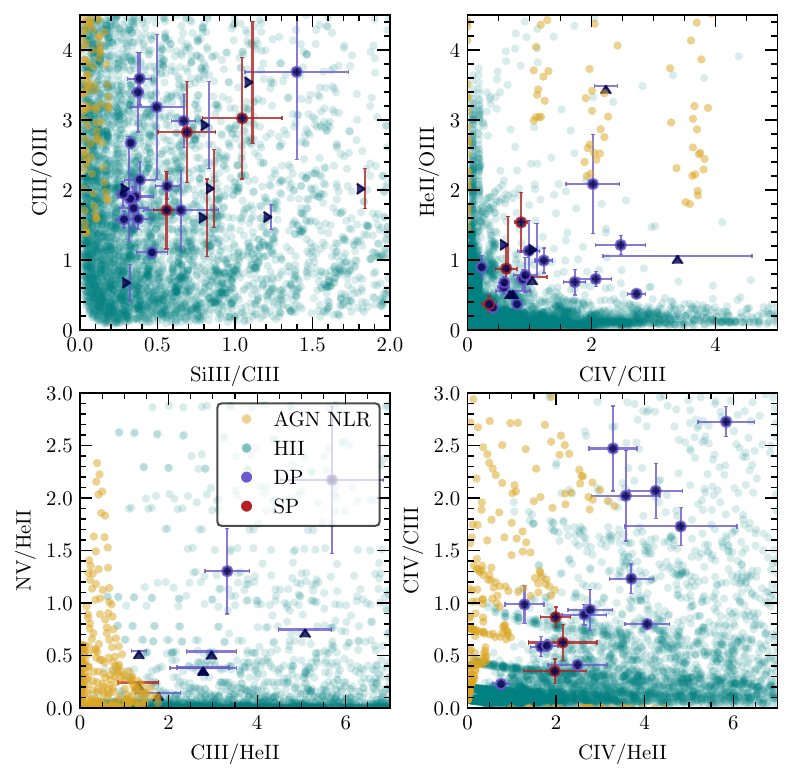"}
	\caption{Rest-UV emission line ratios for spectra with verified, highly significant metal emission line detections.  Circular points indicate detections; vertical (horizontal) triangular points indicate upper bounds in the ordinate (abscissa).  Sources with single- and double-peaked \lya are shown in red and blue, respectively.  Model grids for AGN NLR \citep{feltre16} and star-forming \ion{H}{2} regions \citep{gutkin16} are shown for comparison.  All measured ratios are consistent with photoionization by young stars.}
	\label{fig:line_ratios}
\end{figure}

The most commonly detected metal emission lines in our sample are the \civ, \oiii, and \ciii doublets, as well as the \heii\footnote{Though He is not strictly classified as a metal, we use the term ``metal'' lines loosely here to refer to everything except for H lines.} line.  Due to their high ionization potentials, all of these lines, especially \civ and \heii, are associated with extreme ionization conditions and low metallicities not seen in typical \ion{H}{2} regions in the local universe \citep{stark14,berg21}.  To verify that these lines arise from photoionization by young stars rather than AGN activity, we compared the measured line ratios in the strongest emitters against photoionization model grids produced by \cite{gutkin16} and AGN narrow-line region (NLR) grids produced by \cite{feltre16}.  We find no cases in which the measured line ratios require AGN to be explained (i.e. are consistent with AGN NLR models but not with \ion{H}{2} region models).  On the other hand, all measured line ratios are consistent with the predictions of \cite{gutkin16} for ionization by young stars, and in the \ciii/\oiii and \siiii/\ciii ratios we find excellent agreement with measurements of low-redshift metal-poor dwarf galaxies by \cite{garnett95}, strengthening the case for emission from low-metallicity, star-forming gas.  The ratios shown in Figure~\ref{fig:line_ratios} are the same as those used by \cite{gutkin16}, selected on the basis of their sensitivity to gas metallicity, ionization parameter, dust depletion factor, C/O abundance, neutral hydrogen density, and upper-mass cut-off of the stellar initial mass function.  Measured ratios are not corrected for dust attenuation, though the impact is expected to be minimal given the proximity in wavelength of all lines studied and the generally low dust attenuation in \lya-emitting galaxies \citep{reddy16}.

\newpage

\section{High-z Double-peaked profiles}
\label{app:highz_dp}

\begin{figure*}[ht!]
	\centering
	\includegraphics[width=0.9\textwidth]{"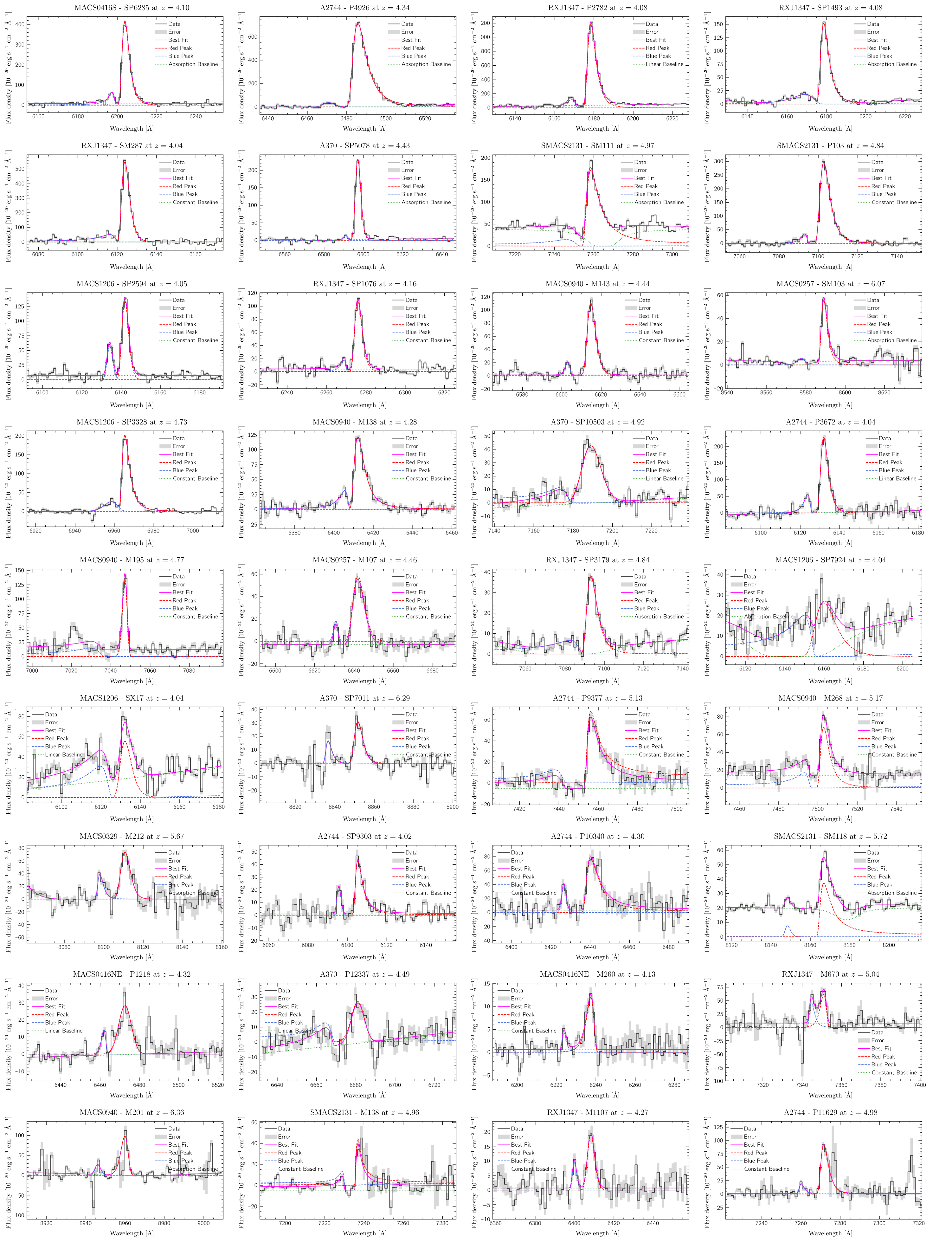"}
	\caption{Double-peaked \lya profiles with $ z \geq 4 $ (showing only those with blue peak SNR $ > 5 $).}
	\label{fig:highz_dp_profiles}
\end{figure*}

\newpage

\section{Emission Line EWs used in Stellar Population + ISM Fitting}
\label{app:ew_table}

\begin{deluxetable*}{ccccccc}[ht!]
	\tablehead{
\colhead{ID} & \colhead{Cluster} & \colhead{\civ} & \colhead{\heii} & \colhead{\oiii} & \colhead{\ciii} & \colhead{\siiii}
}
\startdata
P2861 & MACS1206 & $14.13 \pm 2.09$ & $5.92 \pm 0.76$ & $6.07 \pm 0.53$ & $8.06 \pm 0.91$ & $100.00^{\wedge}$ \\
P4556 & RXJ1347 & $3.88 \pm 0.42$ & $1.87 \pm 0.20$ & $5.60 \pm 0.56$ & $6.68 \pm 0.90$ & $2.56 \pm 0.63$ \\
P477 & MACS0257 & $17.25 \pm 1.96$ & $4.64 \pm 1.31$ & $7.92 \pm 1.06$ & $18.36 \pm 4.07$ & $11.01^{\wedge}$ \\
SM127 & MACS2214 & $1.60 \pm 0.41$ & $0.75 \pm 0.20$ & $4.62 \pm 0.91$ & $4.92 \pm 1.12$ & $100.00^{\wedge}$ \\
SP113 & MACS0257 & $6.93 \pm 0.47$ & $1.75 \pm 0.23$ & $3.53 \pm 0.31$ & $3.82 \pm 0.46$ & $1.11 \pm 0.49$ \\
SP1172 & MACS0416NE & $0.86 \pm 0.07$ & $0.53 \pm 0.06$ & $1.08 \pm 0.07$ & $1.54 \pm 0.09$ & $0.88 \pm 0.11$ \\
SP1200 & MACS0416NE & $0.77 \pm 0.06$ & $0.33 \pm 0.06$ & $0.63 \pm 0.05$ & $0.98 \pm 0.07$ & $0.37 \pm 0.09$ \\
SP1515 & MACS0416NE & $3.68 \pm 0.19$ & $1.03 \pm 0.12$ & $4.01 \pm 0.19$ & $6.46 \pm 0.27$ & $2.04 \pm 0.23$ \\
SP1569 & MACS0416NE & $4.95 \pm 0.48$ & $1.65 \pm 0.43$ & $8.80 \pm 1.05$ & $20.99 \pm 3.05$ & $5.07 \pm 0.87$ \\
SP1580 & MACS0416NE & $4.00 \pm 0.08$ & $0.69 \pm 0.07$ & $2.08 \pm 0.06$ & $1.49 \pm 0.07$ & $0.67 \pm 0.14$ \\
SP482 & MACS0257 & $0.41 \pm 0.04$ & $0.32 \pm 0.11$ & $0.53 \pm 0.08$ & $0.46 \pm 0.07$ & $0.10^{\wedge}$ \\
SP5369 & A370 & $6.77 \pm 1.00$ & $3.64 \pm 0.58$ & $7.82 \pm 1.94$ & $7.77 \pm 1.65$ & $100.00^{\wedge}$ \\
SP5523 & A370 & $1.24 \pm 0.05$ & $0.35 \pm 0.05$ & $0.52 \pm 0.06$ & $1.05 \pm 0.11$ & $0.16^{\wedge}$ \\
SP8031 & MACS0416S & $3.65 \pm 0.29$ & $1.82 \pm 0.29$ & $2.31 \pm 0.40$ & $7.49 \pm 0.91$ & $100.00^{\wedge}$ \\
SP8822 & MACS0416S & $0.40 \pm 0.08$ & $0.55 \pm 0.08$ & $1.00 \pm 0.10$ & $2.33 \pm 0.13$ & $1.60 \pm 0.17$ \\
\enddata
\caption{Rest-frame emission line EWs in $ \mathrm{\AA} $ used to fit stellar population/ISM models in Section~\ref{subsection:ew_factors}. Upper limits are indicated by $^{\wedge}$. All \civ EWs represent lower bounds due to the fact that this is a resonance doublet and so has an escape fraction $ < 1$.}
\end{deluxetable*}

\FloatBarrier
	
\bibliography{/home/james/Documents/my_papers/references/master}

\end{document}